\documentclass[useAMS,usenatbib]{mn2e}
\usepackage[dvips]{graphicx}
\usepackage{amssymb}

\newcommand{\kms}{{\rm \,km\,s^{-1}}}
\newcommand{\mpc}{{\rm Mpc}}
\newcommand{\beq}{\begin{equation}}
\newcommand{\eeq}{\end{equation}}
\newcommand{\petroratio}{{{\mathcal{R} }_P}}

\def\dPA{\delta{\rm PA}}
\def\rp{{R_{50}}}
\def\rel{R_{\rm el}}
\def\bare{{\epsilon}}
\def\A4{{\cal A}_4}

\begin{document}

\title[Isophotal Shapes of E/S0 galaxies]
{Isophotal shapes of Elliptical/S0 Galaxies from the Sloan Digital Sky Survey}
\author[C. N. Hao et al.]{
C. N. Hao,$^{1,2,3}$\thanks{E-mail:hcn@bao.ac.cn}
S. Mao,$^{4,3}$
Z. G. Deng,$^{5}$
X. Y. Xia,$^{3}$
Hong Wu$^{1}$\\
$^{1}$National Astronomical Observatories,
                Chinese Academy of Sciences, A20 Datun Road, Beijing 100012,
                China\\
$^{2}$Graduate School of the Chinese Academy of Sciences,
                 Beijing 100049, China\\
$^{3}$Department of Physics, Tianjin Normal University,
        Tianjin 300074, China\\
$^{4}$Jodrell Bank Observatory, University of Manchester,
          Macclesfield, Cheshire SK11 9DL, UK\\
$^{5}$College of Physical Science, Graduate School of
        the Chinese Academy of Sciences, Beijing 100049, China\\}
\maketitle

\label{firstpage}

\begin{abstract}
We revisit the shapes of isophotes for elliptical (E) and lenticular (S0) 
galaxies by studying
847 nearby early-type galaxies selected from the Sloan Digital 
Sky Survey Data Release 4 with velocity dispersions above $200\kms$.
The IRAF task {\tt ellipse} was used to
derive the deviations of the isophotes from pure ellipses (Fourier coefficients 
$a_3/a$ and $a_4/a$), position angles and ellipticities as a
function of radius. We show the statistical distributions of the $a_3/a$
and $a_4/a$ parameters as a function of velocity dispersion,
magnitude, and colour. The $a_4/a$ parameter is correlated
with the ellipticity and absolute Petrosian magnitude of galaxies.  No 
significant correlation was found between the $a_4/a$ parameter with colour
and velocity dispersion.  A cross-correlation between the FIRST
survey and the SDSS data reveals a lack of powerful radio emitters in discy
E/S0s, as previously found by Bender et al. 
We also find that boxy E/S0s favor denser environments while
discy E/S0s favor more isolated environments.
The median values of changes in the ellipticity and position
angle between one and one and a half Petrosian half-light radii
in the isophotes are about $-$0.023 and $1.61^\circ$
respectively. The average change in the position angle is much
larger, about $4.12^\circ$, due to an extended tail.
The change in ellipticity is weakly correlated with the ellipticity itself,
with an increasing ellipticity for galaxies with higher ellipticity as
the radius increases.
The isophote parameters for the 847 galaxies are available online.
\end{abstract}
\begin{keywords}
galaxies: elliptical and lenticular, CD -- galaxies: photometry --
galaxies: structure
\end{keywords}

\section{Introduction}

The morphologies of galaxies carry important
information on how they form, evolve and how they are supported dynamically.
For example, disc galaxies are highly flattened systems that
are rotationally supported. It implies that angular momentum
must have played a crucial role in determining their properties, such as
sizes and the Tully-Fisher relation (e.g., Mo, Mao \& White 1998 and
references therein).
Similarly, shapes of elliptical (E) and lenticular (S0) galaxies carry 
important information about their formation
history and dynamical state.

The isophotes of elliptical galaxies and S0's can be well fitted by
ellipses to the zeroth order. Small but significant deviations from
perfect ellipses exist. Around 1980's, CCDs became more widely used in
astronomy and they enabled accurate photometric measurements of galaxies.
Several authors studied the isophotal shapes of elliptical galaxies
(Lauer 1985; Bender \& M\"ollernhoff 1987; Jedrzejewski 1987). 
In an influential study, Bender et al. (1988) presented the isophotal
analysis of a sample of 69 nearby bright elliptical galaxies. This study
used CCD photometry in the Cousins $V$, $R$, $I$ under typical
seeing conditions of about 2 arcseconds with a field of view of
3 by 4 arc minutes. Combining CCD data
from other studies, Bender et al. (1989) classified 109 early-type galaxies 
according to the isophotal shapes using Fourier expansions in the polar angle
(see \S\ref{sec:fourier}). It was
found that the most significant non-zero component of the Fourier
analysis is the $a_4$ parameter (corresponding to the $\cos 4\theta$
term). Using the sign of this parameter,  E/S0 galaxies were classified 
into discy ($a_4 > 0$) and boxy ($a_4 < 0$) galaxies.
Bender et al. (1989) showed that this parameter has significant correlations with
the radio and X-ray properties in elliptical galaxies. 
It is now increasingly clear that early-type galaxies can be subdivided into two
classes: discy ellipticals tend to be fainter, rotationally supported,
lack X-ray and radio activities and have power-law nuclear light profiles 
while boxy ellipticals tend to be 
brighter,  supported by random motions, have significant X-ray and radio activities 
and cored nuclear profiles (Ferrarese et al. 1994; van den Bosch et al. 1994; 
Lauer et al. 1995; Faber et al. 1997; Rest et al. 2001;
Lauer et al. 2005). In addition, the shapes of elliptical galaxies may also
be correlated with their ages (Ryden, Forbes \& Terlevich 2001).
Pellegrini (1999, 2005) updated the
analysis of the correlation between galaxy shapes and X-ray properties for 
early-type galaxies using ROSAT and Chandra data. Pasquali et al. (2006) 
found, among other things, that the percentages of discy and boxy ellipticals  at $z\sim 1$
are similar to the local values, and their characteristic shapes
also follow the same correlations of local ellipticals, with an
exception of the $a_3/a$ parameter which appears to be larger than that
for the local counterparts.

The ellipticities and position angles in ellipticals vary as a 
function of distance from the galaxy centre. Di Tullio (1978, 1979) found
that for isolated ellipticals, on average, the ellipticity decreases as
a function of radius, while for ellipticals in clusters and groups, the
ellipticity either increases, decreases or peaks as a function of radius.
This environmental dependence of isophotal shape changes, if confirmed,
would imply that external environments have significant
effects on the inner properties of galaxies. 

There have been many theoretical interpretations for the origins of discy and 
boxy ellipticals. For example, Naab, Burkert \& Hernquist (1999, also Nabb \& Burkert 2003 and Naab, Khochfar \& Burkert 2006) 
concluded that equal-mass mergers of two disk galaxies tend to produce boxy 
ellipticals while non-equal mergers produce discy ellipticals. Khochfar \& Burkert 
(2005) used semi-analytical 
simulations to study the origin of the discy and boxy ellipticals. They
concluded that the isophotal shapes of merger remnants depend not only on
the mass ratio of the last major merger but also on the morphology
of their progenitors and the subsequent gas infall. These theoretical
studies and comparisons with high-redshift objects (Pasquali et al. 2006) all used the observational 
constraints from the limited sample of Bender et al. (1989).
Therefore, it is necessary to revisit the properties of isophotal shapes and their
relations with other galactic properties in early-type galaxies using
a larger sample and images with excellent qualities.

The statistics of $a_4/a$ parameter are important for a variety of
applications, for example gravitational lensing. Deviations from the pure
ellipses can induce higher-order singularities such as butterflies
and swallowtails, which can in turn produce sextuplet and octuplet images (Keeton, Mao \& Witt
2000; Evans \& Witt 2001). Evans \& Witt (2001) estimated that perhaps
$\sim 1$ per cent of galaxy-scale lenses are sextuplet and octuplet imaged
systems. The study of a large and homogeneous sample of
ellipticals and S0's is important for providing sufficient statistics of
the shapes of E/S0 galaxies, which dominate the lensing cross-section on galaxy scales.

In this paper, we use the Sloan Digital Sky Survey (SDSS) Data Release 4 (DR4) 
data of 847 nearby E/S0 galaxies to 
study the distributions of their isophotal parameters and the relations
of isophotal shapes with other galactic properties.
The SDSS survey provides high-quality photometry in five colours
under typical seeings of about 1.5 arcseconds over a large area of the sky.
In \S 2, we discuss how we select our sample of elliptical
and S0 galaxies, outline the data reduction procedures and how we estimate
the parameters that describe the shapes of E/S0 galaxies. The size of our sample
is about a factor of 8 larger than that used by Bender
et al. (1989). This large sample size allows us to consider more carefully 
how the shapes
of galaxies are correlated with a variety of parameters, such as their
velocity dispersions, luminosities, and colours. 
We also consider how their shapes are correlated with
the environments. These results are presented in \S 3. We discuss our results in the 
last section (\S 4). Throughout this paper, we adopt a cosmology with
a matter density parameter $\Omega_{\rm m}=0.3$, a cosmological constant
$\Omega_{\rm \Lambda}=0.7$ and
a Hubble constant of $H_{\rm 0}=70\,{\rm km \, s^{-1} Mpc^{-1}}$.

\section{Sample Selection and Data Analysis \label{sec:sample}}

\subsection{Sample selection \label{sec:samsel}}

Our sample objects are selected from the SDSS DR4 photometric catalogue. 
We select all objects according to the
following criteria: (1) Their redshifts must be smaller than 0.05; this
is to ensure sufficient spatial resolutions to resolve these galaxies.
(2) The velocity dispersions cover a range of 200 $\kms$ to 420 $\kms$.
The latter is approximately the maximum velocity dispersion that can be measured reliably.
(3) The target galaxies should not be saturated or located at the
edge of the corrected frame (Stoughton et al. 2002). In total, there are
1102 galaxies satisfying the above three criteria. However, 11 of these 1102 objects 
are unavailable from the SDSS DR4 Data Archive Server. Furthermore, to ensure 
the early-type galaxy morphology of our sample, we visually examined all images 
and find 916 ($\approx 83$ per cent) of them are E/S0 galaxies and not 
contaminated by companion galaxies or bright stars. To reduce the impact
of dust on the photometry, objects with visible dust lanes are also
excluded from our sample. As we will estimate the intensity-weighted
Fourier coefficients and ellipticity, $\bare$, (see \ref{subsec:estimate}) 
within a region from twice the seeing radius ($2\, r_{\rm s}$) to 1.5 times 
Petrosian half-light radius ($1.5 \rp$), objects with 
$2 r_{\rm s} >= 1.5\rp$ are excluded from the following 
analyses. The $r$-band seeing radius for each object is the effective
PSF width determined at the center of each frame by the SDSS pipeline (Stoughton et al. 2002).
However, in most cases, our objects are not located at the centres.
Stoughton et al. (2002) pointed out that the PSF FWHM can
vary by up to 15\% from one side of a CCD to the other, even in the 
absence of atmospheric inhomogeneities. We tested the effect on
isophotal parameters by enlarging the effective PSF width by 10\%
and found that the median differences for the weighted means of the 
$a_3/a$, $a_4/a$, $b_3/a$, $b_4/a$
and ellipticity are 0.072, 0.15, 0.020, 0.030
and 2 times their corresponding errors respectively. The relatively large
difference in ellipticity ($\sim 2\sigma$) between these two data sets is probably due to the 
monotonic ellipticity radial profiles for a large fraction of our sample
objects and the small (formal) errors ($\sigma$) in the ellipticity.

For our sample objects, the redshifts are low and thus should be 
corrected for the Local Group infall. We adopted the Local Group relative
redshifts provided by Blanton et al. (2005) to calculate the luminosity 
distances.
There are three objects without such information and another two with negative
redshifts. These five objects are excluded from the following analyses. 
So our final catalogue includes 847 E/S0s.

\subsection{Data reduction}

\subsubsection{Image reduction}

We use the SDSS $r-$band images in this paper to perform isophotal analyses.
The corrected frame obtained from SDSS is bias-subtracted, flat-fielded,
and corrected for cosmic rays and pixel defects (Stoughton et al. 2002). Before
surface photometry, sky background fitting and subtraction are
performed (Wu et al. 2005). To do this, we detect all objects by SExtractor 
(Bertin \& Arnouts 1996) and mask them to generate a background-only image. 
The sky background is obtained by fitting this image and then subtracted from
the corrected frame. After background subtraction, the corrected frame is
trimmed to a 501x501 frame which centres on the target.

SExtractor was then run on the trimmed frame to generate a ``SEGMENTATION" image 
(See documents for SExtractor), which identifies all objects in the frame.
A mask image with all detected objects except the galaxy of interest
flagged can then be obtained from the ``SEGMENTATION" image.
We carefully examined all the mask
images, and found that SExtractor performs satisfactorily for masking
neighbour objects in most cases. For a few (five) rare cases that SExtractor
did not work well, we carefully tuned the parameters in SExtractor to create 
good mask images. Surface photometry is performed on the trimmed image
using the IRAF task {\tt ellipse} (Jedrzejewski 1987) 
with the masked areas excluded from the fitting. 
This task requires some initial guesses for the geometric centre, ellipticity 
and position angle for the isophotes; these guesses are obtained by running 
SExtractor on the masked image. When running {\tt ellipse} on the targets, 
we allow the geometric centre, ellipticity and position angle to vary freely and assume a 
logarithmic step of 0.1 in radius along the semi-major axis.

\subsubsection{Isophotal shapes from Fourier analysis \label{sec:fourier} }

In the literature, the isophotes of elliptical galaxies have been measured
in two different ways, which we outline below.
\begin{enumerate}
\item The first method first finds the isophote for a given surface brightness 
$I_0$ (by interpolation, for example). The isophote can be described by
$R(\theta)$, where $(R, \theta)$ are the polar coordinates centred on
the centre of an ellipse.
In general, the isophote is not a
  perfect ellipse. The {\it deviation} of the isophote from a perfect
  ellipse can be expanded in Fourier series in the polar angle
	\begin{equation}
		\delta R(\theta) = R(\theta)-\rel(\theta)= a_0 + \sum (a_n \cos n \theta + b_n \sin n \theta)
\label{eq:rtheta}
	\end{equation}
where $\rel(\theta)$ describes the best-fitting ellipse,
and $a_0$ is the average deviation averaged
over the polar angle. The first five coefficients ($a_0$, 
$a_1$, $b_1$, $a_2$ and $b_2$) should be zero within the errors as they
are found from fitting. The lowest order significant deviations are
$a_3, b_3, a_4, b_4$ etc.
\item In the second method, an ellipse is drawn to approximately match an isophote. The
  intensity along the ellipse is then expanded in Fourier series
\begin{equation}
		I (\theta) = I_0 + \sum (A_n \cos n \theta + B_n \sin n \theta)\label{eq:Itheta}
	\end{equation}
	where $I_0$ is the intensity averaged over the ellipse, and $A_n$ and $B_n$ are
	the higher order Fourier coefficients.
	If an isophote is a perfect ellipse, then all the
        coefficients, $(A_n, B_n), n=1, \cdot\cdot\cdot, \infty$ will be exactly zero.
\end{enumerate}
The study by Bender et al. (1988) adopts the first approach while
the IRAF task {\tt ellipse}\footnote{IRAF is distributed by the National Optical Astronomy Observatories,
    which are operated by the Association of Universities for Research
    in Astronomy, Inc., under cooperative agreement with the National
    Science Foundation.} adopts the second method. {\tt Ellipse}
outputs $A_n$ and $B_n$ divided by the semi-major axis length, $a$, and the
local gradient. In the Appendix we show that this essentially gives $a_n/a$ and
$b_n/a$ parameters, as used by Bender et al. (1988, 1989). So our results
can be directly compared with those of Bender et al. (1988, 1989).

\subsection{Estimates of parameters\label{subsec:estimate}}

The characteristic size (Petrosian half-light radius here), velocity dispersion and 
apparent Petrosian magnitude of galaxies are derived from the DR4 photometric 
catalogue. SDSS defines the Petrosian ratio $\petroratio$ at a radius $R$ from
the centre of an object to be the ratio of the local surface brightness in an
annulus at $R$ to the mean surface brightness within $R$
(Blanton et al. 2001; Stoughton et al. 2002):
\beq
\petroratio (R) \equiv
\frac{\int_{0.8 R}^{1.25 R} dR' 2\pi R' I(R') /
[\pi(1.25^2 - 0.8^2) R^2] }
{\int_0^R dR' 2\pi R' I(R') / (\pi R^2)},
\eeq
where $I(R)$ is the azimuthally averaged surface brightness profile.  The
Petrosian radius $R_{\rm P}$ is defined as the radius at which
$\petroratio(R_{\rm P})$
equals 0.2. The Petrosian flux in any band is then defined to be the flux
within two Petrosian radii:
\beq
F_{\rm P} \equiv \int_0^{2 R_{\rm P}} 2\pi R'dR' I(R').
\eeq
The Petrosian half-light radius $R_{50}$ is the radius within which 
50 per cent of the Petrosian flux is contained. The Petrosian quantities
provided by SDSS are derived using circular apertures. For galaxies
with large ellipticities, this will give a different $R_{50}$ compared with that
using elliptical apertures. However, the adoption of circular apertures
will not affect our results statistically as there are few very elliptical galaxies
in our sample (Fig.~\ref{fig:eHist}). To be more quantitative, we calculated the Petrosian
$R_{50}$ using elliptical apertures for some sample objects. Even for the most
elliptical galaxies in our sample, the difference 
between circular and elliptical Petrosian $R_{50}$ is only up to $\sim$ 13\%.
Such a small change in $R_{50}$ will not change our results significantly (see below).

We derived the absolute Petrosian magnitude from the apparent Petrosian 
magnitude and the extinction ($A$) in each filter provided by SDSS DR4 
photometric catalogue by $M=m-5\,{\rm log}(D_{\rm L}/10{\rm pc})-A$.
As the redshift of our sample is low ($< 0.05$) and the range of the 
redshift is small (from 0.01 to 0.05), no k-correction is applied here.
Using the k-corrections (corrected to $z=0.0$) provided by Blanton et al. 
(2005; see also Blanton et al. 2003), we find that the median and maximum 
k-corrections in the $r$-band for our sample objects are
0.04 and 0.06 mag respectively, and hence this should not affect
our results significantly.

As the velocity dispersion provided by SDSS is that in the
3$\arcsec$ fiber aperture, we correct it to a standard circular aperture 
defined to be one-eighth Petrosian half-light radius similar to Bernardi et al. (2003a).
It should be noted that Bender et al. (1988, 1989) used the effective radius 
(half-light radius) from de Vaucouleurs law fitting as the characteristic size 
of a galaxy. We use the Petrosian half-light radius since it is a well 
defined quantity, while in contrast, the 
derivation of effective radius depends on how well the de Vaucouleurs law
fits the surface brightness profile, the range of radius used in the fit
(see Bender et al. 1989 and references therein) and how well the sky 
background is subtracted in the photometry.

Isophotal surface photometry can provide the structural parameters,
such as Fourier coefficients ($a_3/a$, $a_4/a$, $b_3/a$, $b_4/a$), ellipticity 
and position angle (PA) in each elliptic
annulus, where $a$ is the length of the semi-major axis of the fitted ellipse.

In this paper, we mainly study isophotes using quantities averaged over 
intensities or those at one or one and a half Petrosian half-light radii.
Note that the radial profiles of isophotal parameters we used here are along the 
geometric mean radius, $\sqrt{ab}$, where $a$ and $b$ are the lengths of
semi-major and semi-minor axes of a fitted ellipse.
To be more specific, the characteristic values of Fourier coefficients and 
ellipticity are calculated by weighting them with the intensity (counts) in the 
isophote and inversely with their RMS errors over a region of 2$r_{\rm s}$ to 1.5$\rp$.
This is similar to that of
Bender et al. (1988), who used the mean $a_3/a$ and $a_4/a$ between twice the
seeing radius and 1.5 effective radii. Note, however, that Bender et al. (1988, 1989) used
a different definition of characteristic ellipticity -- They adopted the peaked
value for a peaked ellipticity radial profile and the ellipticity at the 
effective radius in case of a monotonically increasing or decreasing 
ellipticity radial profile.
The changes in $a_3/a$, $a_4/a$, ellipticity, PA and the centres of ellipses 
between $\rp$ and $1.5\rp$ are also estimated. The standard deviations of 
these quantities are calculated by error propagation. As expected, 
fainter objects have larger errors than the brighter ones.
 All the data are available online
\footnote{http://www.jb.man.ac.uk/\~\,smao/isophote.html}.
 
The surface brightness profiles of E/S0's are often fitted with the
S\'ersic profile (S\'ersic 1968). From the fitting, it is
possible to obtain the s\'ersic index $n$ and a s\'ersic effective radius 
within which half of the
light is contained. Blanton et al. (2005) also gave the s\'ersic index $n$ and 
$R_{50}$ in their tables. We compared the SDSS Petrosian $R_{50}$ and 
the s\'ersic $R_{50}$ given by Blanton et al. (2005) and found that they differ 
by 14 per cent on average. Blanton et al. (2005) argued that their s\'ersic 
$R_{50}$ is underestimated
by $\sim$ 10 per cent for large sizes and high s\'ersic indices. We thus
enlarged the Petrosian half-light radius $R_{50}$ by 25 per cent and 
re-computed the isophotal parameters
and compared them with those derived by SDSS Petrosian $R_{50}$. The
average differences between these two sets of data are 0.016, 0.023, 0.0045, 0.0073
and 1.76 times the standard deviations for the weighted means of the 
$a_3/a$, $a_4/a$, $b_3/a$, $b_4/a$ and ellipticity; the median difference 
between these two data sets for 
the above five parameters are respectively 0, 0, 0.0066$\sigma$, 0.0049$\sigma$ 
and 0.67$\sigma$, so our results are quite robust with respective to small
uncertainties in $R_{50}$. The relatively large difference in ellipticity 
(average $\sim 1.76\sigma$ and median $\sim 0.67\sigma$)
is possibly due to the monotonic ellipticity radial profiles for a large 
fraction of our sample objects and the small (formal) errors ($\sigma$) in 
ellipticity.

\section{Results}

\subsection{Sample properties}

\begin{figure*}
 \begin{minipage}[t]{0.4\linewidth}
   \centerline{\includegraphics[width=0.99\textwidth]{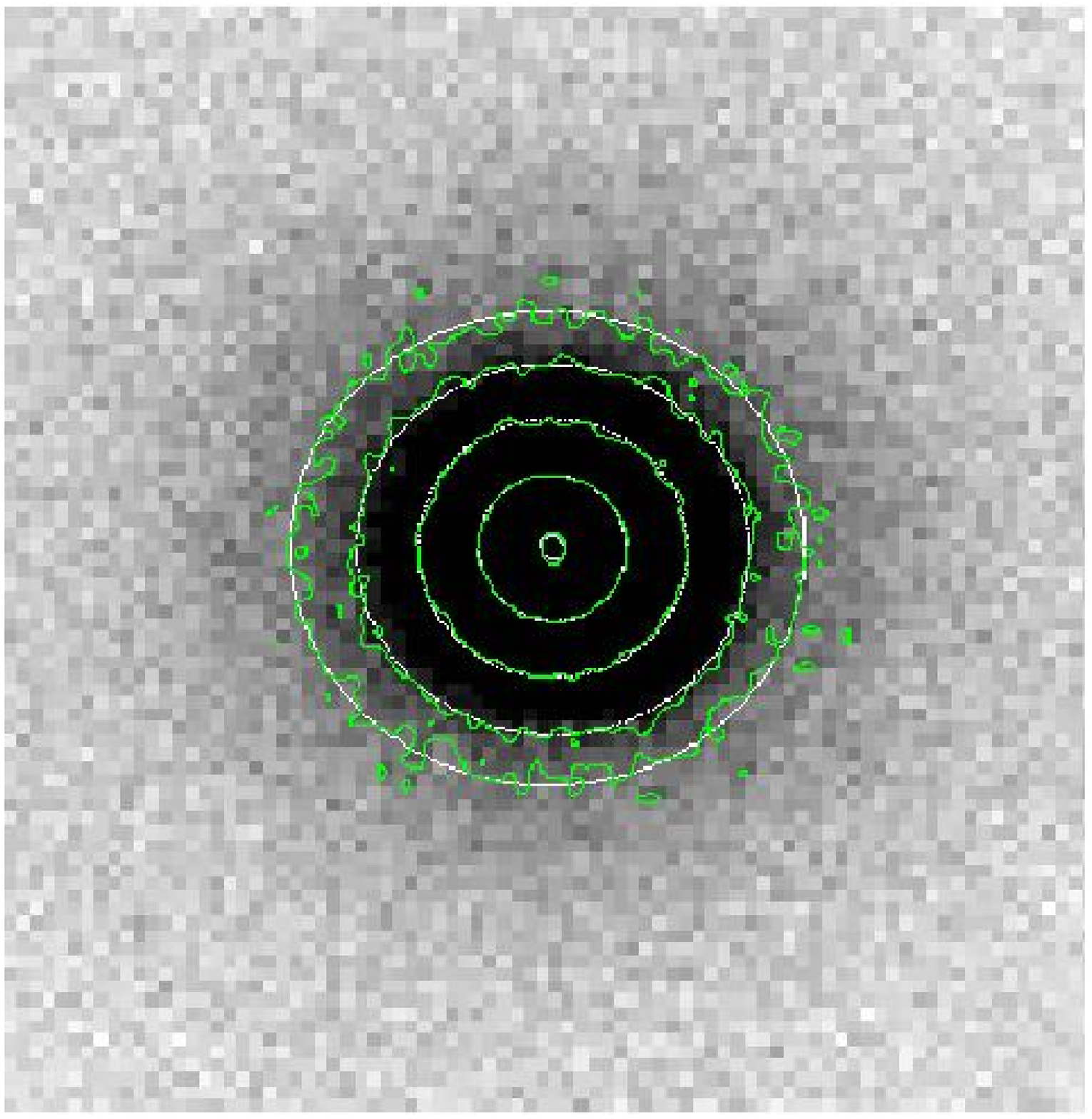}}
   \end{minipage}%
 \begin{minipage}[t]{0.4\linewidth}
    \centerline{\includegraphics[width=0.99\textwidth]{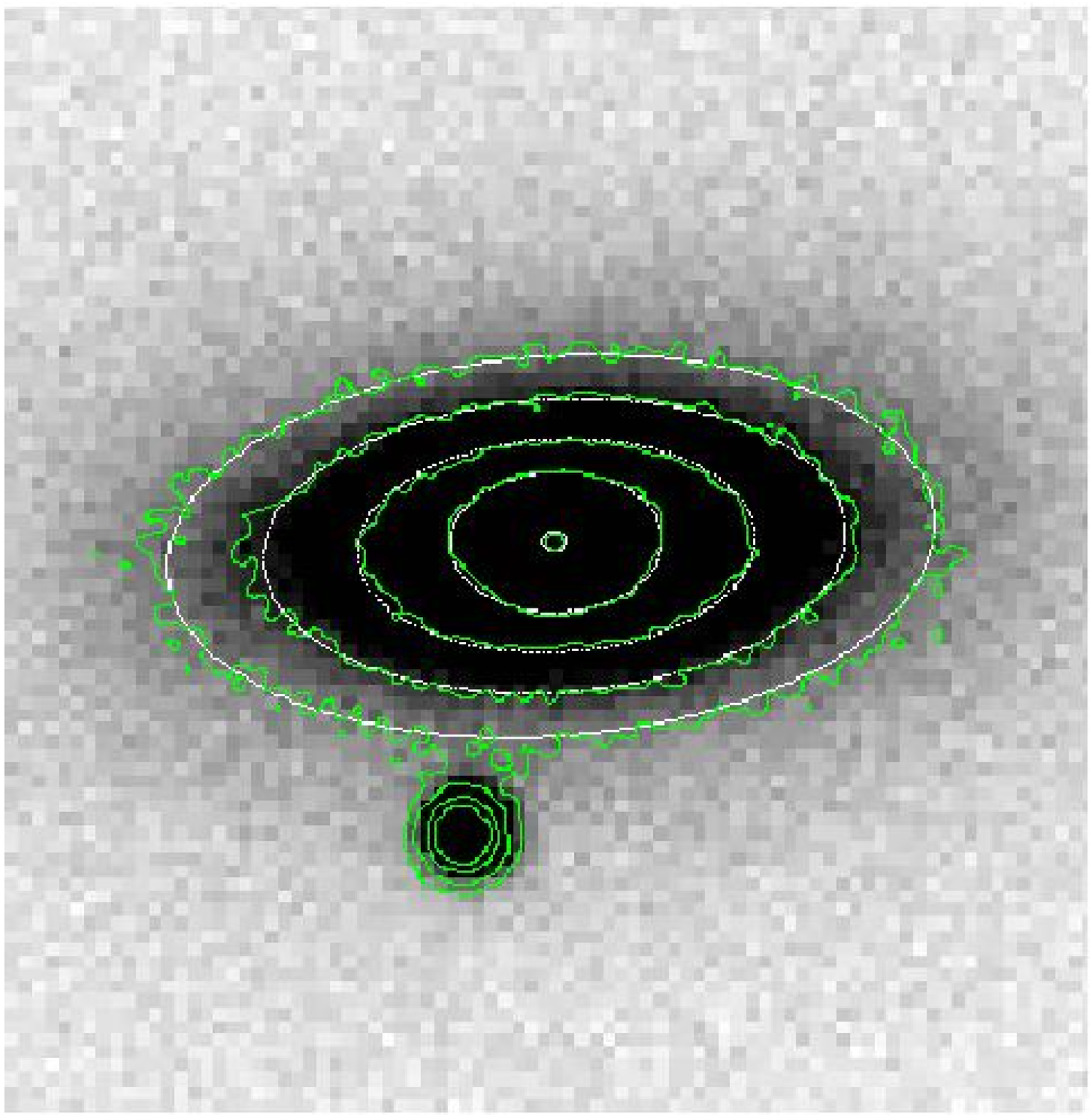}}
   \end{minipage}%
\\
 \begin{minipage}[t]{0.4\linewidth}
   \centerline{\includegraphics[width=0.99\textwidth]{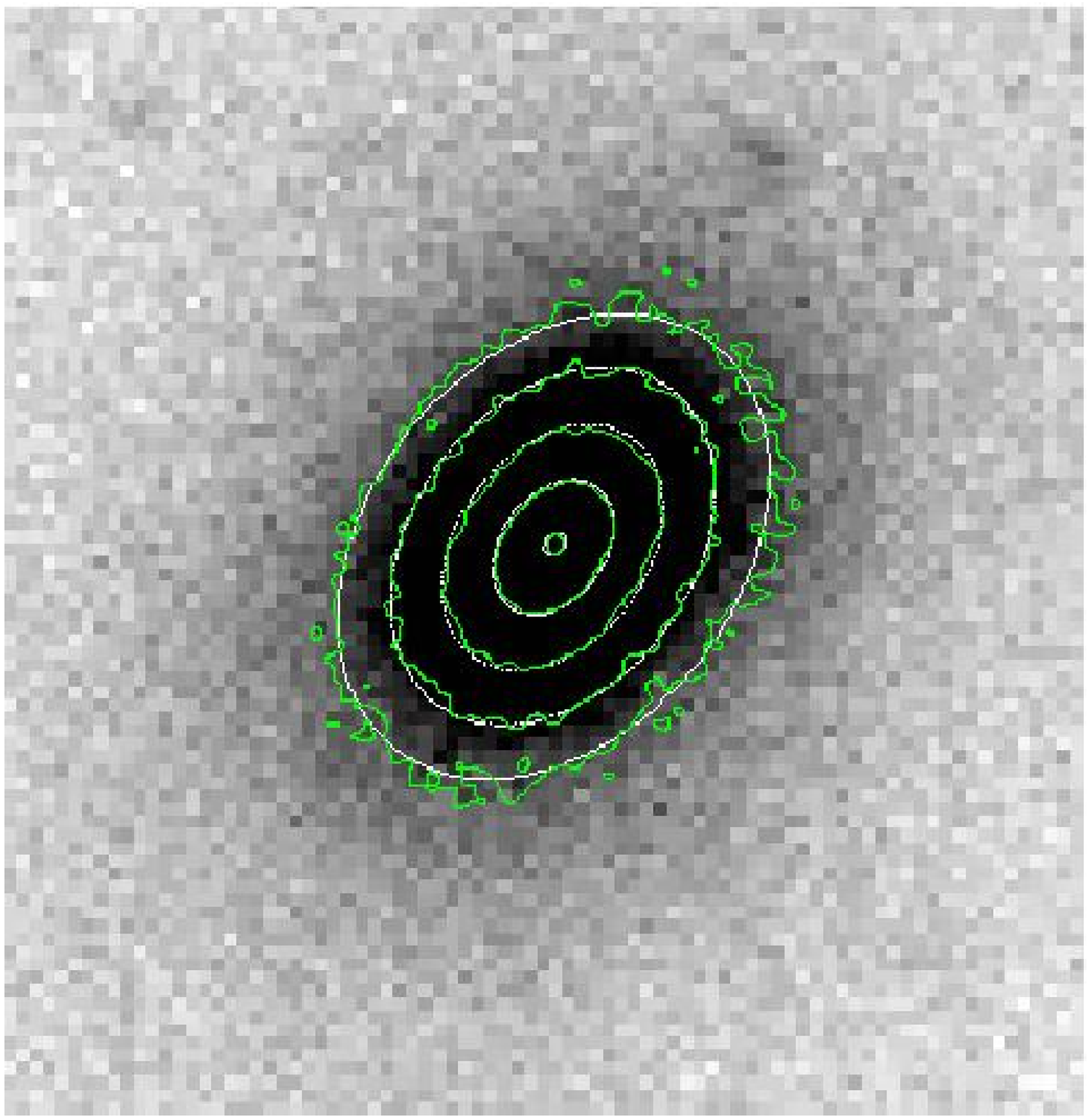}}
   \end{minipage}%
 \begin{minipage}[t]{0.4\linewidth}
   \centerline{\includegraphics[width=0.99\textwidth]{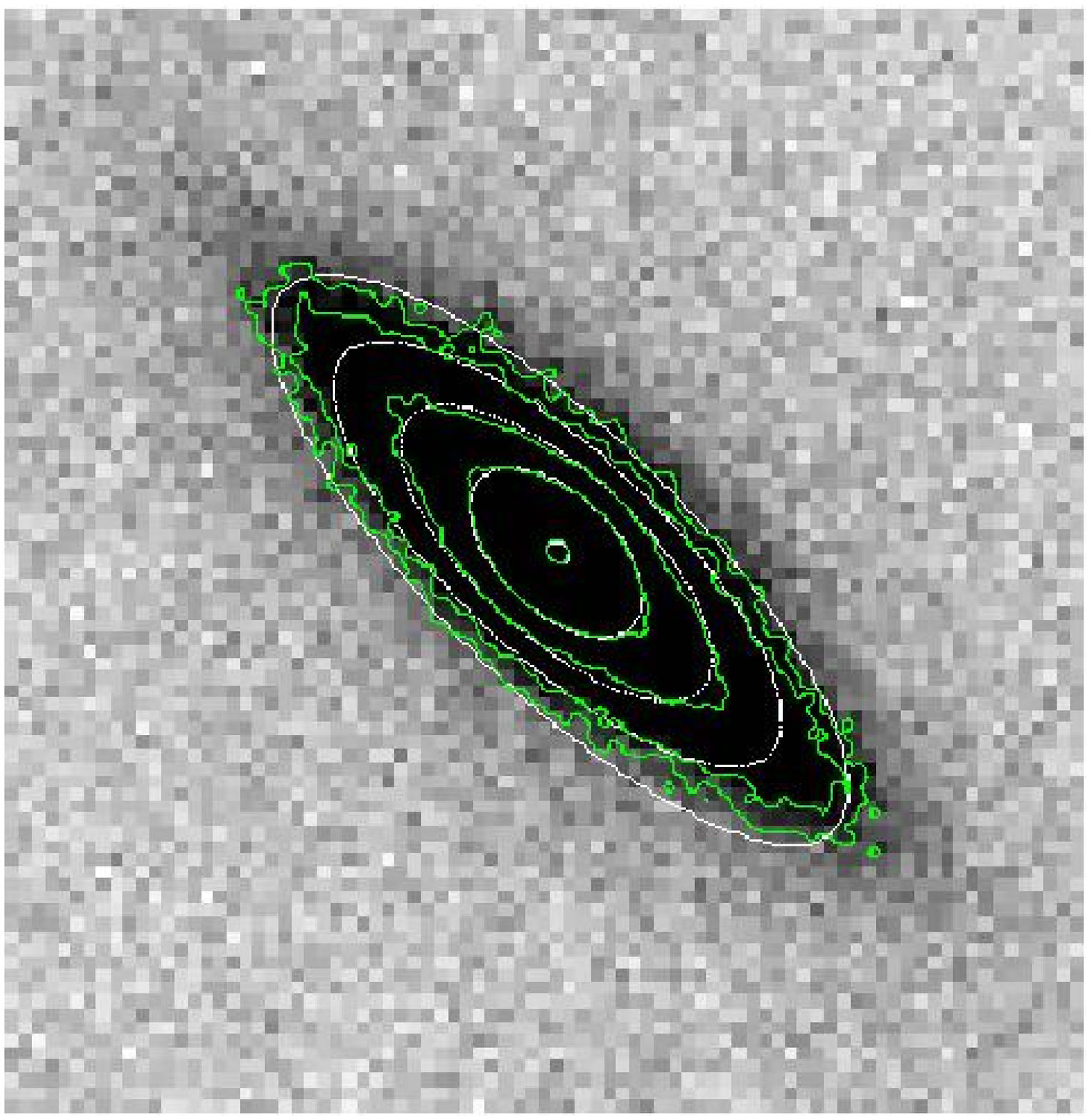}}
   \end{minipage}%
\caption{Four examples of our sample galaxies. The top two panels
show two ellipticals that are well fitted by ellipses within noises. The
ellipticities of these two galaxies are respectively 0.03 and 0.46.
The bottom two panels show one boxy (left, $a_4/a=-0.02$) and one discy
(right, $a_4/a=0.035$) elliptical galaxy.
The field size for each image is 34.4$\arcsec$ by 34.4$\arcsec$. Five isophotes are shown
together with the best-fitting ellipses.}
\label{fig:images}
\end{figure*}

Fig.~\ref{fig:images} shows images for four of our sample galaxies.
The top left and right panels show
one nearly round and one very flattened ellipticals; both of which
can be fitted by ellipses within noises. The bottom two panels show one
example of a boxy elliptical and a discy elliptical respectively.
It is clear that ellipticals and S0's can be reasonably fitted by ellipses but
also show significant deviations. Below we first study some basic
properties of our sample.

\begin{figure*}
\centerline{\includegraphics[width=0.6\textwidth]{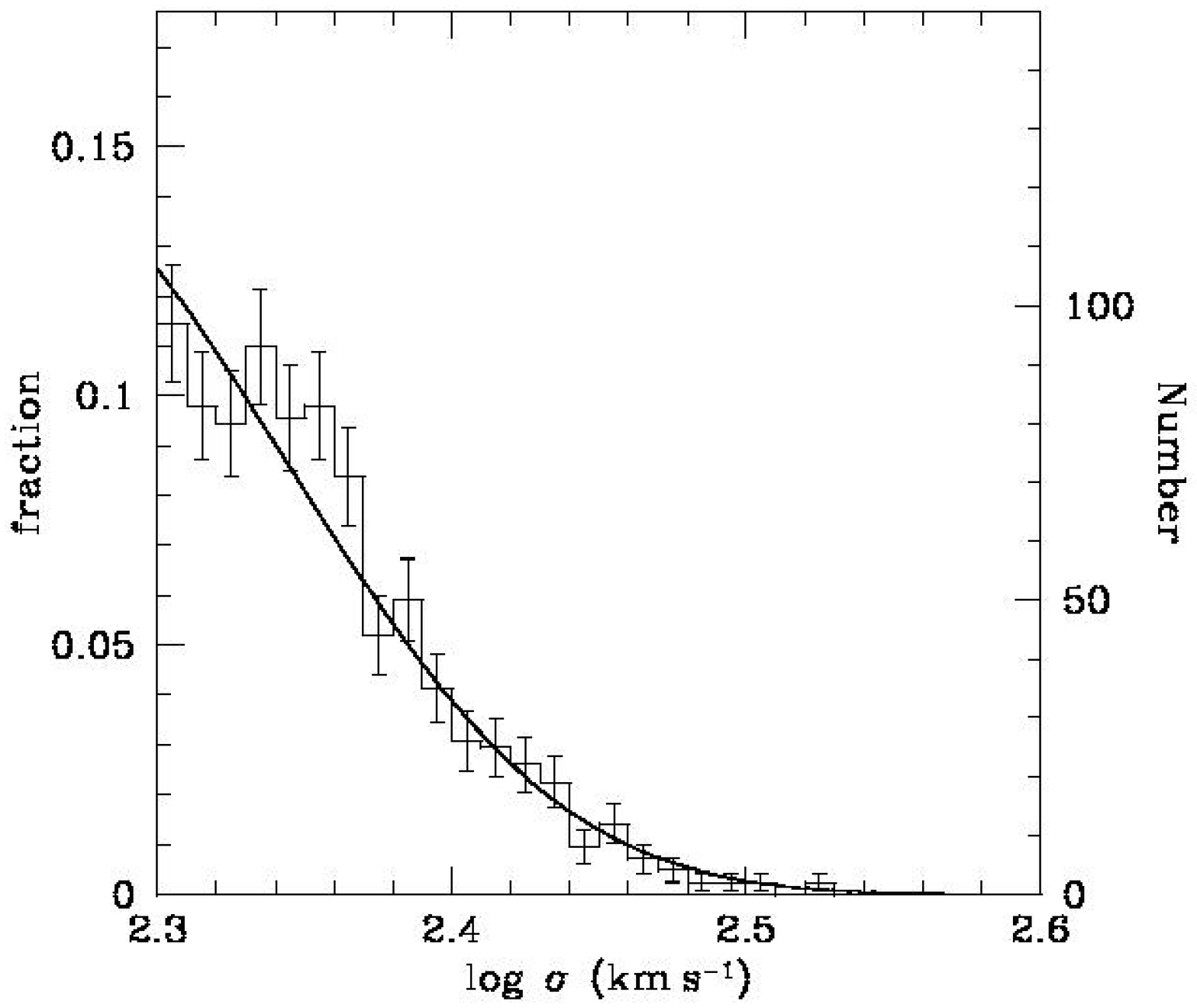}}
\caption{Histogram of velocity dispersion without aperture correction for our
847 galaxies. The solid line is the
velocity dispersion function derived by Sheth et al. (2003) from the SDSS. 
The error bars indicate the Poisson errors. For all histograms in this paper,
the error bars have the same meaning and this will not be stated in the following.
}
\label{fig:sigmaHist}
\end{figure*}

The histogram of velocity dispersion ($\sigma$) for our sample is shown in
Fig.~\ref{fig:sigmaHist}. The solid line shows the velocity dispersion
function derived by Sheth et al. (2003) from an analysis of a 
large sample of galaxies from the SDSS, carefully accounting for the selection
effects and measurement errors. Our histogram matches
the Sheth et al. curve reasonably well except at the small dispersion
end, where our sample seems to show a small deficit but still within
error bars. The reasonable match between 
our sample and the Sheth et al. curve indicates our sample does not
have a strong bias regarding to the velocity dispersion. As the
luminosity is correlated with the velocity dispersion
through the Faber-Jackson relation (Faber \& Jackson 1976; 
Bernardi et al. 2003b), we do not expect our sample to be
significantly biased with respect to the luminosity (or magnitude).

\begin{figure*}
\centerline{\includegraphics[width=0.6\textwidth]{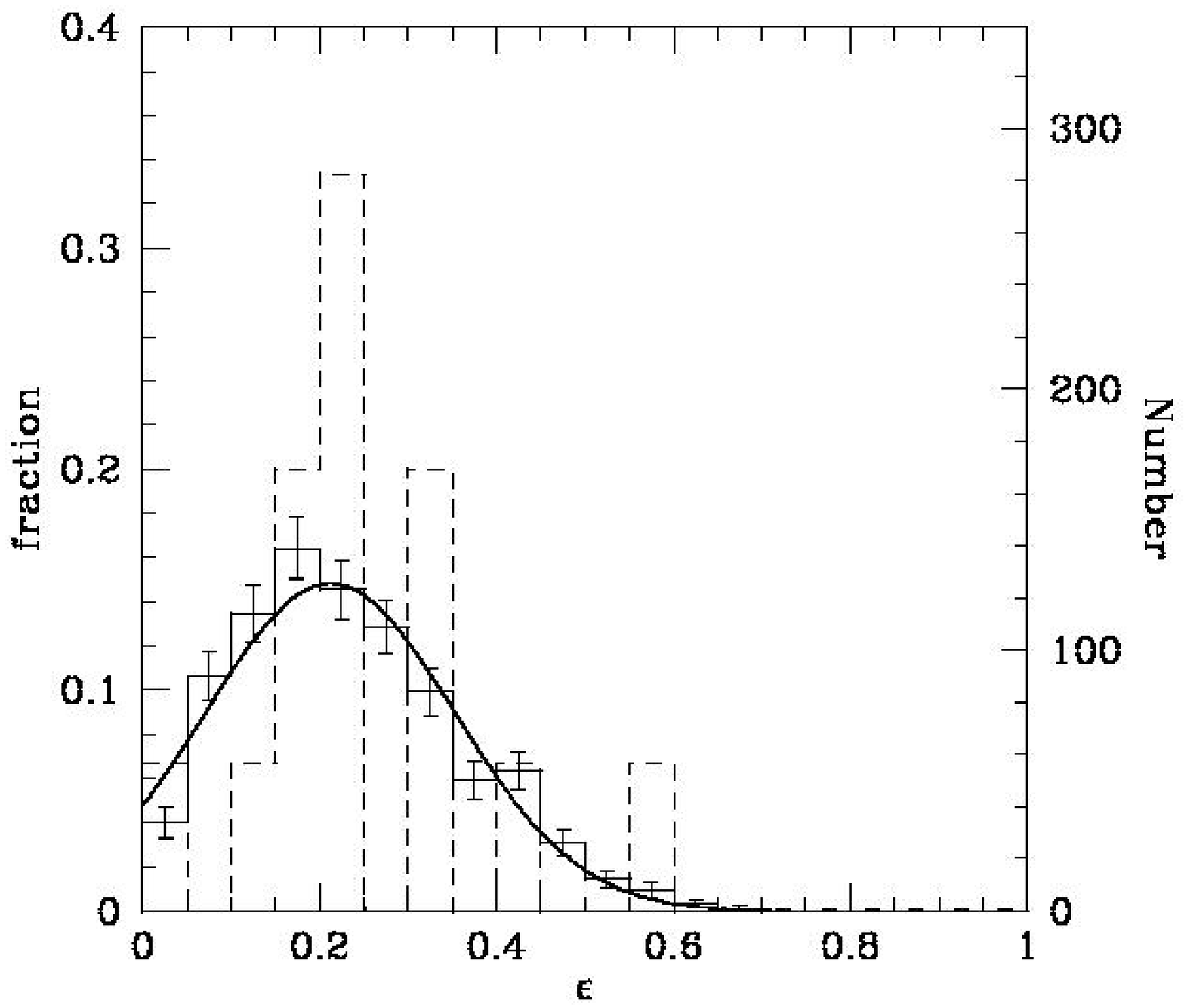}}
\caption{Histogram of ellipticity, $\epsilon$, for our
sample (solid curve). $\epsilon$ is the average ellipticity weighted by 
intensity between twice the seeing and 1.5 Petrosian half-light radii.
The solid line shows an empirical Gaussian fit to the data. The
dashed histogram shows the ellipticity histogram for the 15 lenses
studied by Koopmans et al. (2006).
}
\label{fig:eHist}
\end{figure*}

The ellipticity distribution for our sample is shown in Fig.~\ref{fig:eHist}.
The distribution shows a peak around $\bare \approx 0.2$ and drops off
sharply above $\bare=0.5$ (correspond to an E5 elliptical). The mean and
dispersion in the ellipticity are 0.23 and 0.13 respectively. Our
distribution appears to show smaller ellipticities than the one shown
in Lambas, Maddox \& Loveday (1992) based on the APM survey which uses
photographic plates. Their distribution
peaks around $\epsilon=0.25$ and drops off sharply above
$\epsilon=0.8$. As they adopted a different definition
of ellipticity (their equations 1 and 2), we used similar quantities as
provided by SDSS (See Kuehn \& Ryden 2005 for details) to examine the 
ellipticity distribution. The distributions of these two
ellipticities in the SDSS data are very similar. Therefore, 
the difference between our sample and theirs is 
not due to different definitions of ellipticity; it is not clear 
what causes it. However, our ellipticity distribution is in very
good agreement with Fasano \& Vio (1991), who used CCD images of elliptical
galaxies. The differential probability 
distribution of the ellipticity can be well fitted by a Gaussian: 
\beq
p(\bare) d\bare = {0.052 \over \sqrt{2 \pi} \sigma_{\bare}}
{\rm e}^ {-{(\bare-0.22)^2 \over 2\sigma^2_{\bare}}}\, d{\bare},
\eeq
where the dispersion $\sigma_{\bare}=0.14$.
This fit is shown as the solid curve in
Fig.~\ref{fig:eHist}. The ellipticity distribution (dashed histogram) of
the lens galaxy sample of Koopmans et al. (2006) is shown for comparison in the same Figure.
The Kolmogorov--Smirnov test indicates a 
probability of 58.6 per cent that the two samples are drawn from the same
parent distribution (for more see the discussion).

\begin{figure*}
\centerline{\includegraphics[width=0.6\textwidth]{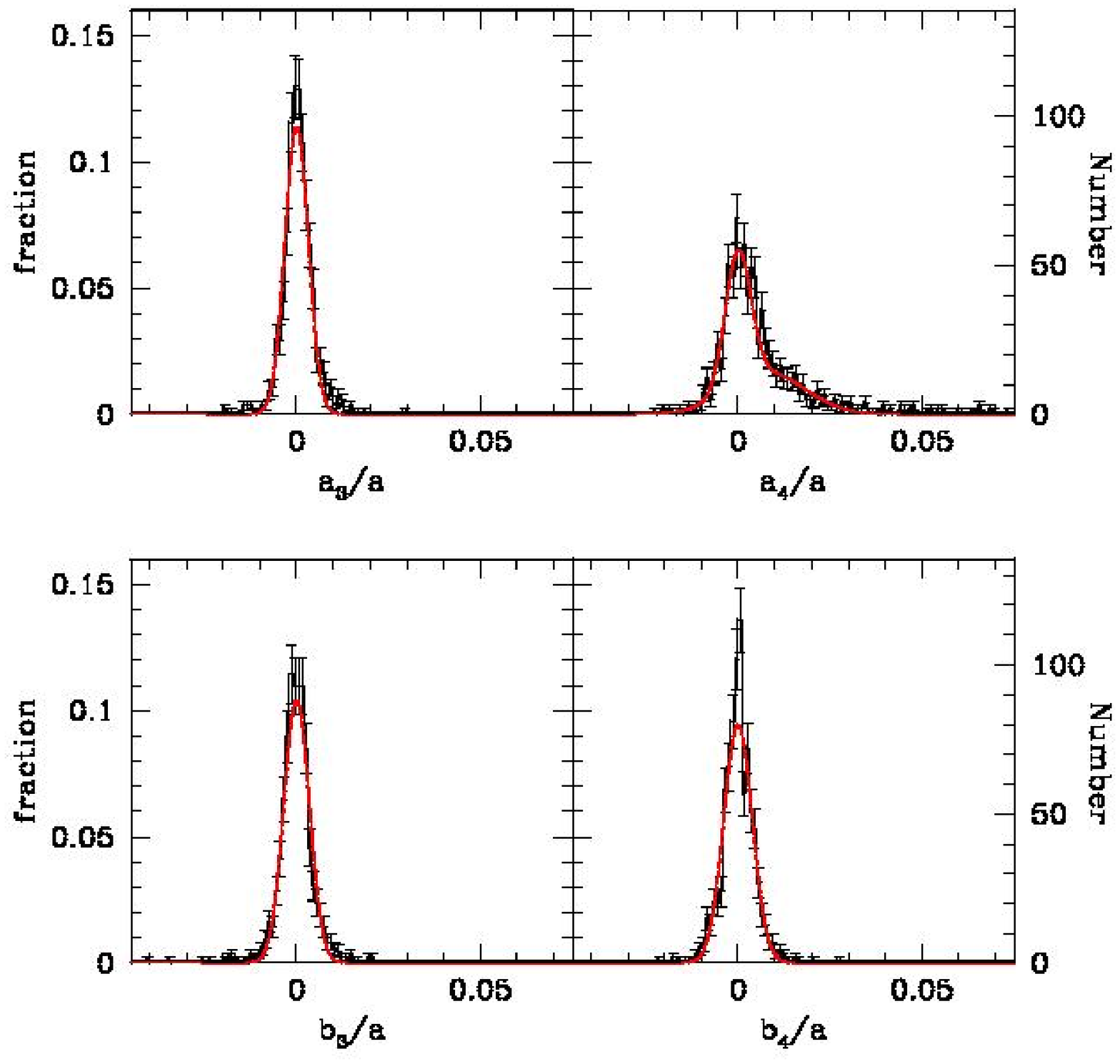}}
\caption{Histograms of $a_3/a$ (upper left), $a_4/a$ (upper
  right), $b_3/a$ (bottom left), and $b_4/a$ (bottom right). The solid
lines are the best-fitting single Gaussian model for $a_3/a$, $b_3/a$, $b_4/a$ and
double Gaussian model for $a_4/a$ (See text).
}
\label{fig:a3a4Hist}
\end{figure*}

The histograms of $a_3/a$, $a_4/a$, $b_3/a$ and $b_4/a$
are shown in Fig.~\ref{fig:a3a4Hist}. The distributions of $a_3/a$, $b_3/a$ and
$b_4/a$ are reasonably fitted by a single standard Gaussian with zero
means and dispersions of 0.0032, 0.0035 and 0.0039 respectively. The histogram for $a_4/a$, on the 
other hand, shows more asymmetries,
particularly toward the positive $a_4/a$ indicating an excess of 
discy E/S0s compared to boxy E/S0s.  The
distribution of $a_4/a$ can be reasonably fitted by the sum of two Gaussian 
functions (The normalized $\chi^2$ per degree of freedom is 0.81, where
the errors are taken as Poisson errors):
\beq
{dP \over d\A4} =
 {4.61 \times 10^{-4} \over \sqrt{2 \pi} \sigma_1} {\rm
   e}^{-{\A4^2 \over 2 \sigma_1^2} } +
 {4.32 \times 10^{-4}
\over \sqrt{2 \pi} \sigma_2} {\rm e}^{-{(\A4-0.0079)^2 \over 2
\sigma_2^2} }
\eeq
where $\A4 \equiv {a_4/a}$, $\sigma_1=0.0035$, and $\sigma_2=0.01$.

\subsection{Correlations with galaxy properties}

\begin{figure*}
\centerline{\includegraphics[width=0.6\textwidth]{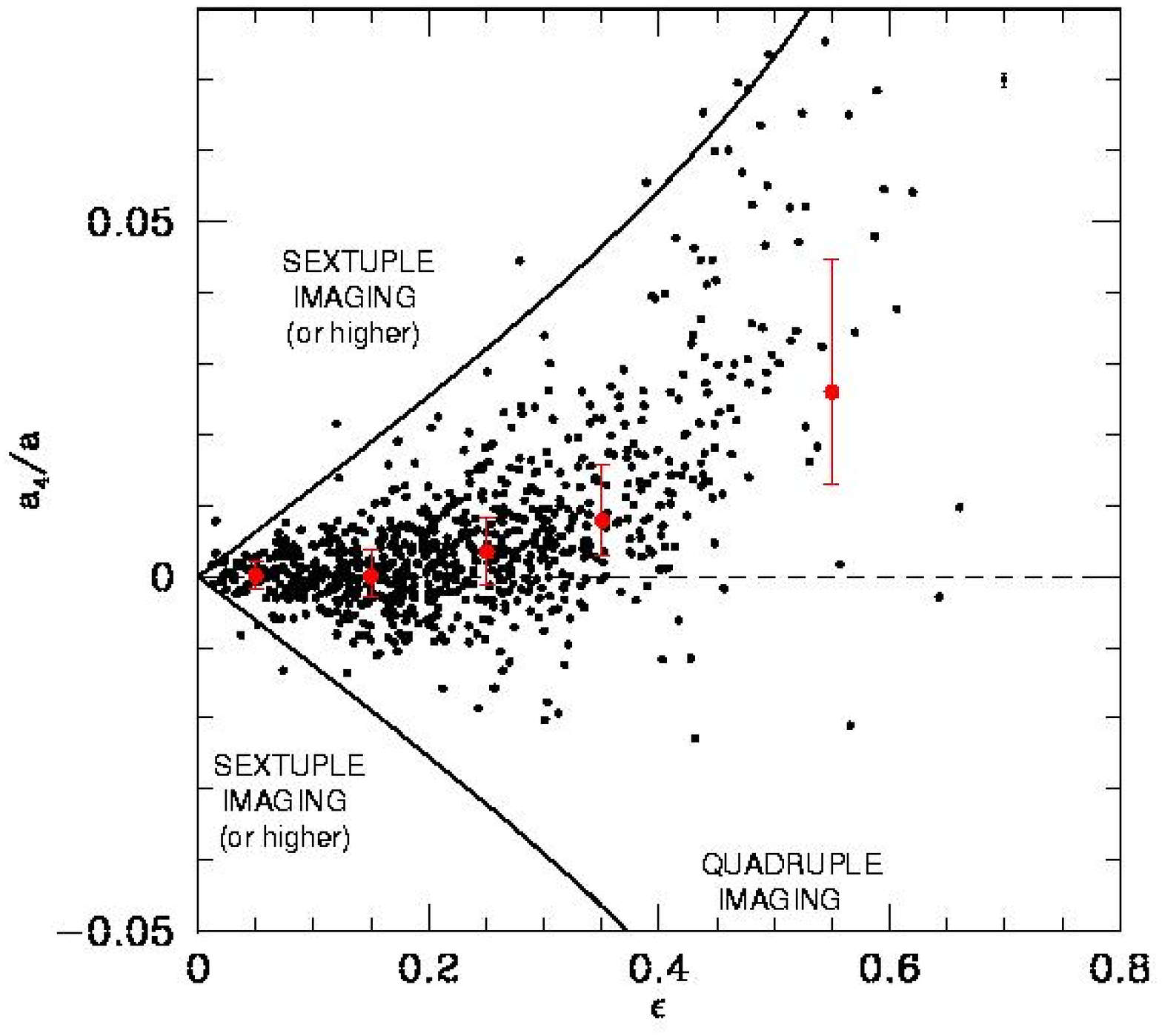}}
\caption{$a_4/a$ as a function of ellipticity. The data points
with error bars are the median, lower (25 per cent) and upper (75 per cent) quartiles
for galaxies in bins of width 0.1 in ellipticity except
the last bin with a width of 0.3 to include all the remaining objects. The
solid curve shows the boundary between quadruple and sextuple lenses
derived by Evans \& Witt (2001). The median error bars are shown at the top right
corner. All error bars shown at the top right corner of the following figures
have the same meaning.}
\label{fig:a4e}
\end{figure*}

Fig.~\ref{fig:a4e} shows how $a_4/a$ varies as a function
of the mean ellipticity, $\bare$. It is clear that when $\bare>0.4$, the
isophotes show significant systematic deviations from perfect ellipses,
with the $a_4/a$ parameter reaching 0.08 or so. As the ellipticity 
is both a function of the intrinsic shape and the viewing angle along
the line of sight, this trend indicates that E/S0s, when viewed edge on,
are more likely to show significant deviations from perfect ellipses. 

\begin{figure*}
\centerline{\includegraphics[width=0.6\textwidth]{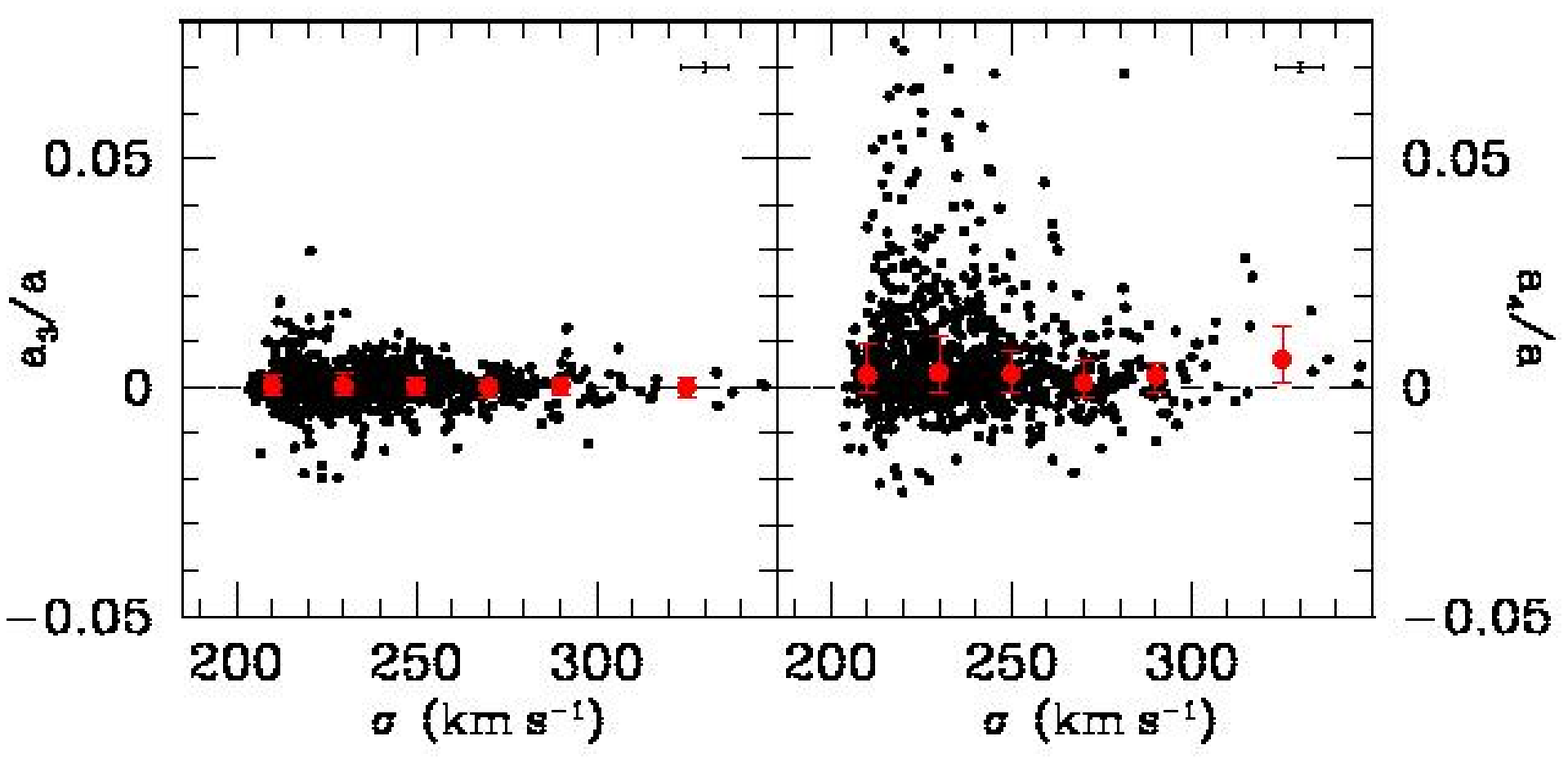}}
\caption{$a_3/a$ (left) and $a_4/a$ (right) as a function of
velocity dispersion. The data points with error bars are the median, lower (25 per cent)
and upper (75 per cent) quartiles for
5 bins separated by 20$\kms$ from $\sigma=200, 300 \kms$. The last bin
has a large bin width of 50$\kms$ to include all the remaining galaxies.}
\label{fig:a3a4sigma}
\end{figure*}

\begin{figure*}
\centerline{\includegraphics[width=0.6\textwidth]{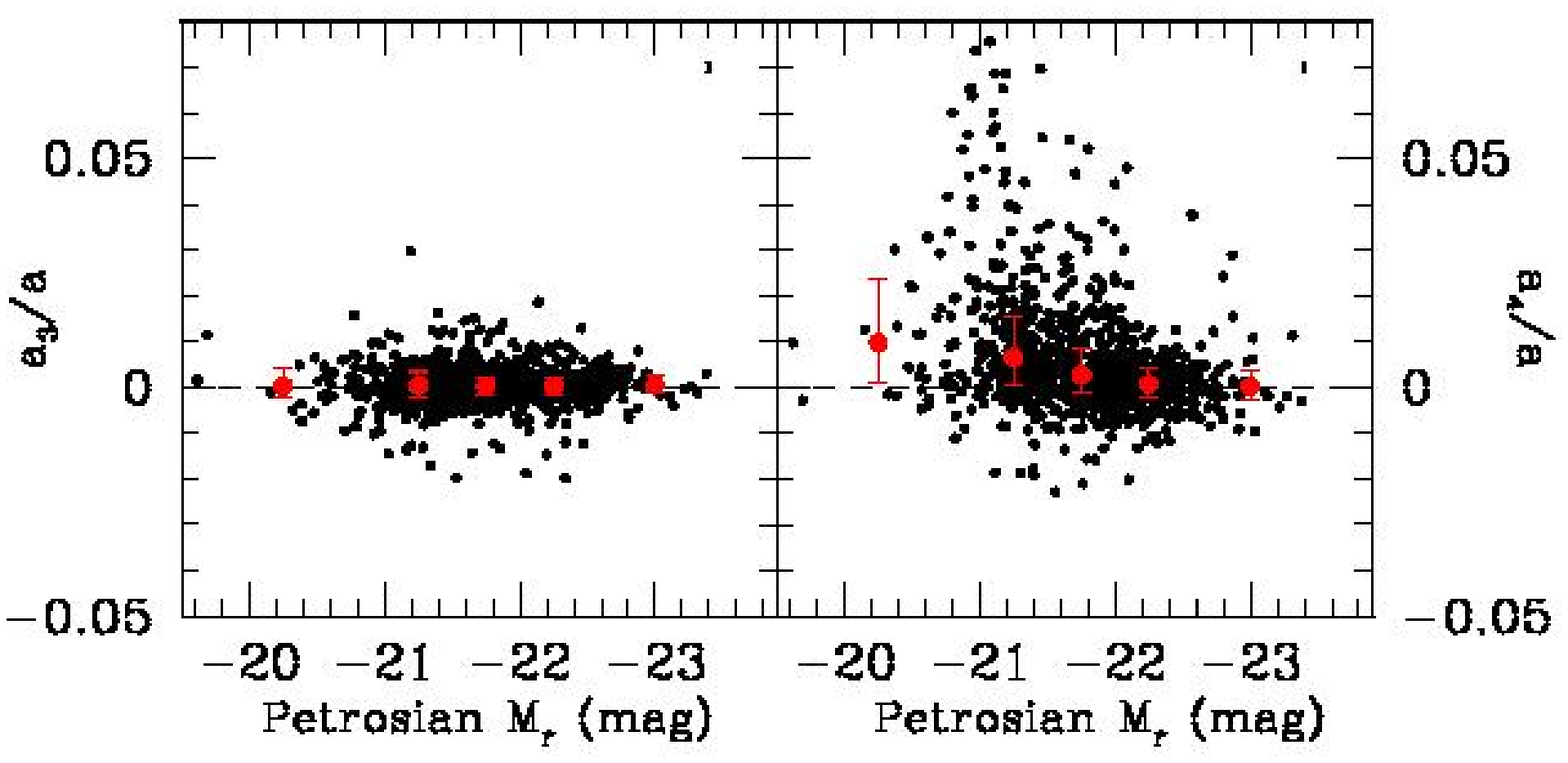}}
\caption{$a_3/a$ (left) and $a_4/a$ (right) as a function of
the absolute magnitude in the $r$-band. 
The median, lower (25 per cent) and
upper (75 per cent) quartiles for all galaxies are shown in 5 bins
with bin width of 0.5 mag except the first and last bin which have a width of
1.5 and 1 mag respectively.}
\label{fig:a3a4mag}
\end{figure*}

Fig.~\ref{fig:a3a4sigma} shows how the $a_3/a$ and $a_4/a$ parameters
vary as a function of the velocity dispersion, $\sigma$.
There is no significant trend for the $a_3/a$ parameter. For the
$a_4/a$ parameter, there seems to be a slight indication that at small
velocity dispersions ($\sigma<250\kms$), there are more E/S0s with
significant non-zero $a_4/a$ parameters, around 0.05. This indicates
a weak trend that more massive ellipticals may have less significant
deviations from perfect ellipses. The trend of $a_3/a$
parameter as a function of $r$-band absolute magnitude is similar to
that of $a_3/a$ versus $\sigma$. The correlation between $a_4/a$ and
the $r-$band absolute magnitude is more significant (Fig.~\ref{fig:a3a4mag}).
The Spearman-Rank Order correlation coefficient
and the probability that no correlation exists for $a_3/a$ vs. $\sigma$,
$a_4/a$ vs. $\sigma$, $a_3/a$ vs. $M_{r}$,
and $a_4/a$ vs. $M_{r}$ are
${\rm r_s}= -0.019,-0.042,0.0054,0.32$ and 
${\rm Prob}=0.58,0.22,0.88,1.84\times10^{-21}$ respectively. Among these correlations,
the relation between $a_4/a$ and $M_{r}$, the $r-$band absolute magnitude, is most
significant; the fraction of boxy galaxies increases with increasing
luminosity. The relations between $a_3/a$, $a_4/a$ and the dynamical
mass, defined as $9 \sigma^2 \rp/G$ (Bender et al. 1989),
are similar to those with the $r-$band absolute magnitude.

Faber et al. (1997) concluded that galaxies with 
$M_V < -22$ have `core' luminosity profile, those
with $M_V > -20.5$ have `power-law' luminosity profile
and at intermediate (with $-22 < M_V < -20.5 $) magnitudes,
`core' and `power-law' coexist. Many studies (e.g., Rest et al. 2001) 
indicated that `core' galaxies are more luminous, slowly rotating and
boxy systems while `power-law' galaxies are less luminous, fast rotating
and discy systems. Therefore, it is interesting to estimate
quantitatively the relative number of 
discy and boxy E/S0s in these magnitude intervals. 
We use the prescription of
$V = g-0.55(g-r)-0.03$ given by Smith et al. (2002) to transform 
the $g$-band magnitude into the $V$-band magnitude for our sample objects. 
Using our isophotal analysis results, we find that
the ratios of the number of discy to boxy E/S0s in these magnitude
intervals ($M_V > -20.5$, $-22 < M_V < -20.5 $ and
$M_V < -22$) are 2.8 (28/10), 2.09 (466/223) and 1.07 (62/58) 
respectively. 
These numbers show that the largest fraction of boxy E/S0s is in the 
brightest subsample, while the number of discy E/S0s increases relative
to boxy E/S0s at the faint end.

\begin{figure*}
\centerline{\includegraphics[width=0.6\textwidth]{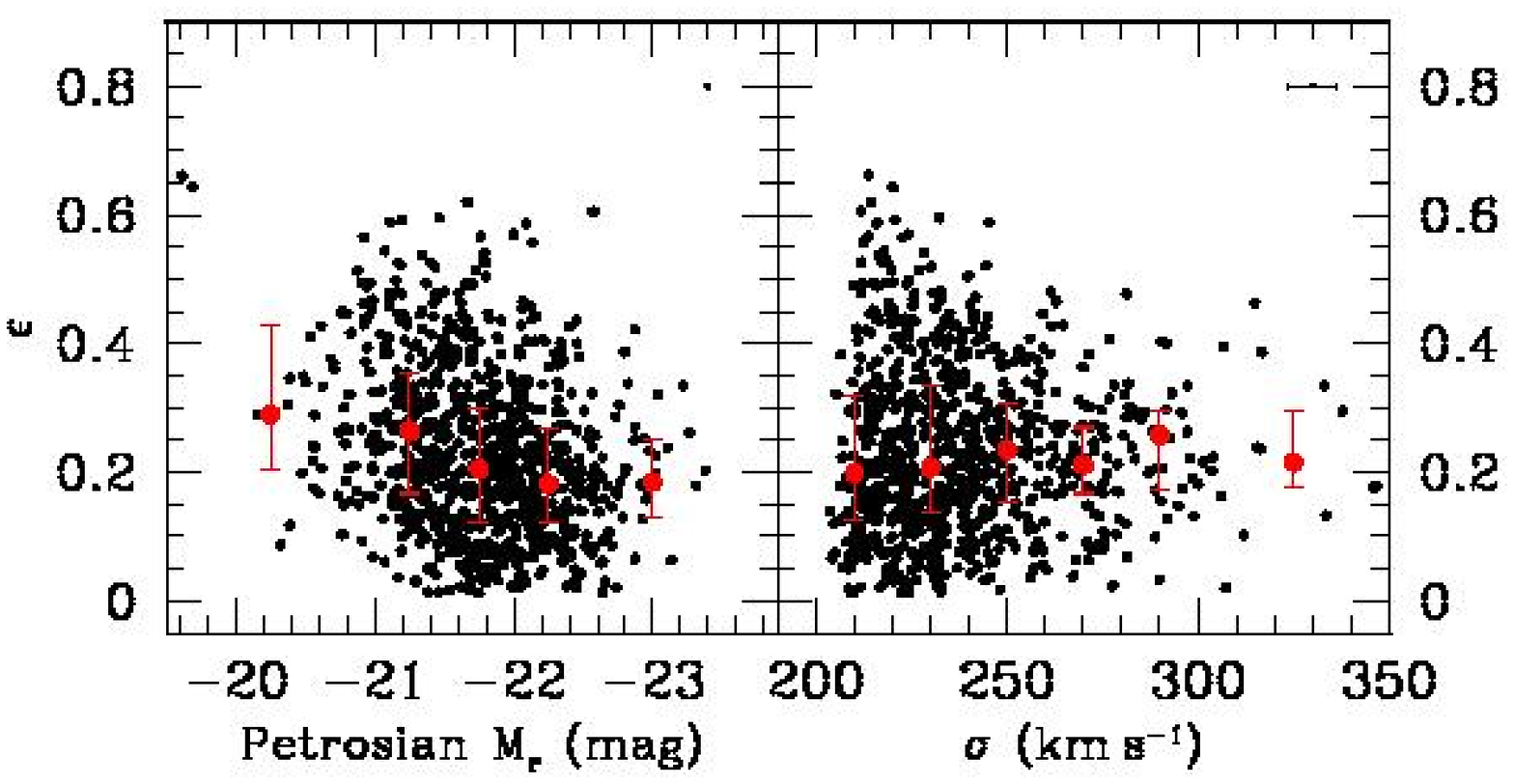}}
\caption{Ellipticity vs. $r$-band absolute magnitude (left) and velocity
dispersion (right) for our sample galaxies. In the left (right) panel, 
the data points with error bars have the same meaning with Fig.~\ref{fig:a3a4mag}
(Fig.~\ref{fig:a3a4sigma}).}
\label{fig:eSigma}
\end{figure*}

Fig.~\ref{fig:eSigma} shows the mean ellipticity, $\bare$, as a function
of the $r-$band absolute magnitude and velocity dispersion.
The data points with error bars show the median values and
the lower (25 per cent) and upper (75 per cent) quartiles
of $\bare$ in bins of absolute magnitude (left panel) and velocity dispersion
(right panel).
In many places in this paper, we quote the median and lower and
upper quartiles as they are less sensitive to outliers (we sometimes
also quote the usual mean and standard deviation value).
As can be seen from the left panel, the ellipticities of E/S0s are correlated
with the absolute magnitudes, although with large scatters: E/S0s
become rounder with increasing
luminosities, which is consistent with Vincent \& Ryden (2005).
In contrast, the ellipticities of E/S0s show no significant correlation
with their velocity dispersions. Quantitatively, a Spearman--Rank Order
correlation analysis
gives a correlation coefficient of 0.26 at $>99.9$ per cent significance level 
for $\bare$ versus $r-$band absolute magnitude, while for $\bare$ versus 
velocity dispersion, the correlation coefficient is 0.035 at a significance 
level of 69.2 per cent. We also find that the mean ellipticity has
a strong correlation with the dynamical mass, with a correlation coefficient
of $-$0.27 at $>99.9$ per cent significance level.

\begin{figure*}
\centerline{\includegraphics[width=0.6\textwidth]{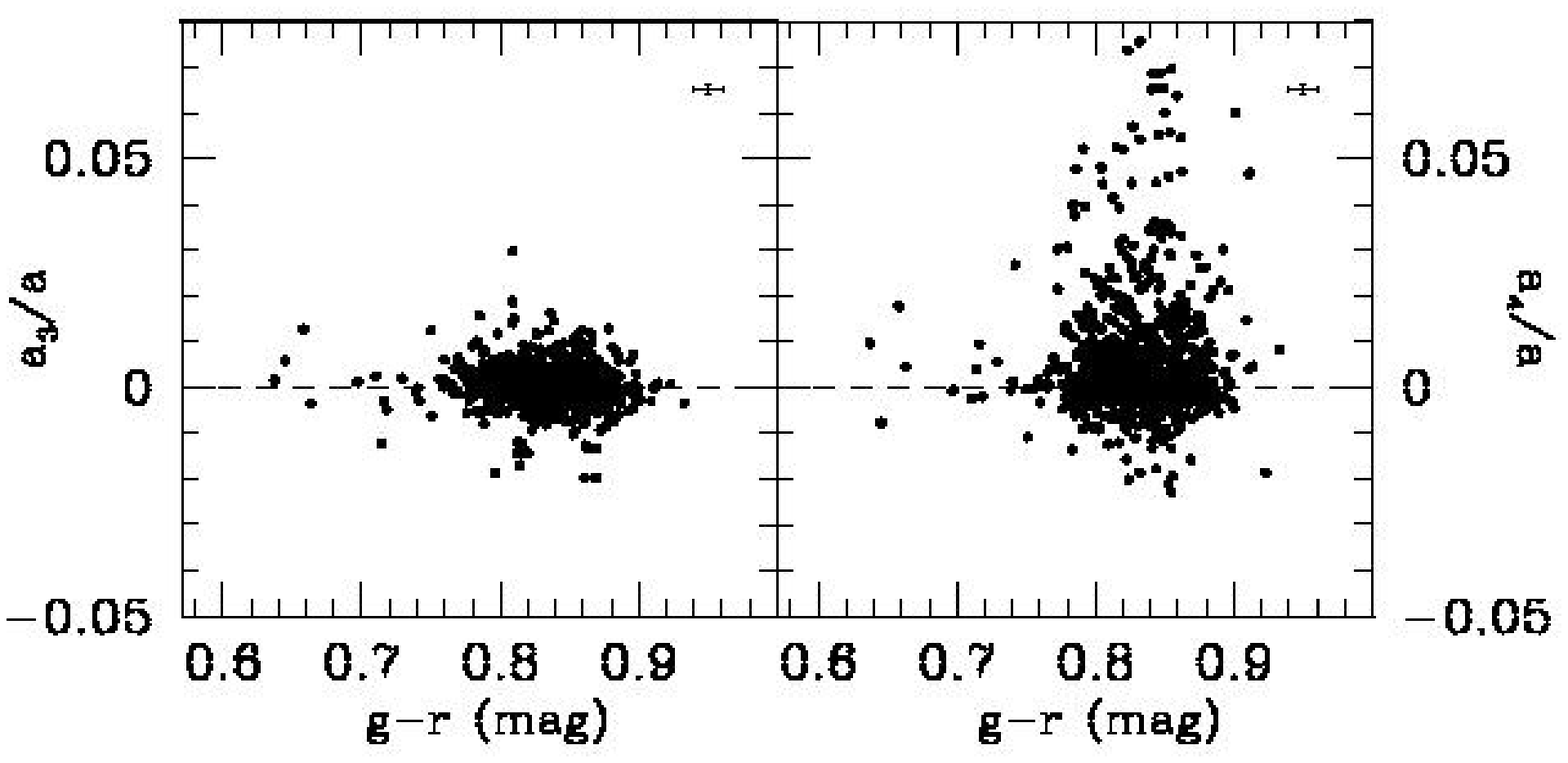}}
\caption{$a_3/a$ (left) and $a_4/a$ (right) parameters as a function of
color.}
\label{fig:a3a4color}
\end{figure*}

\begin{figure*}
\centerline{\includegraphics[width=0.6\textwidth]{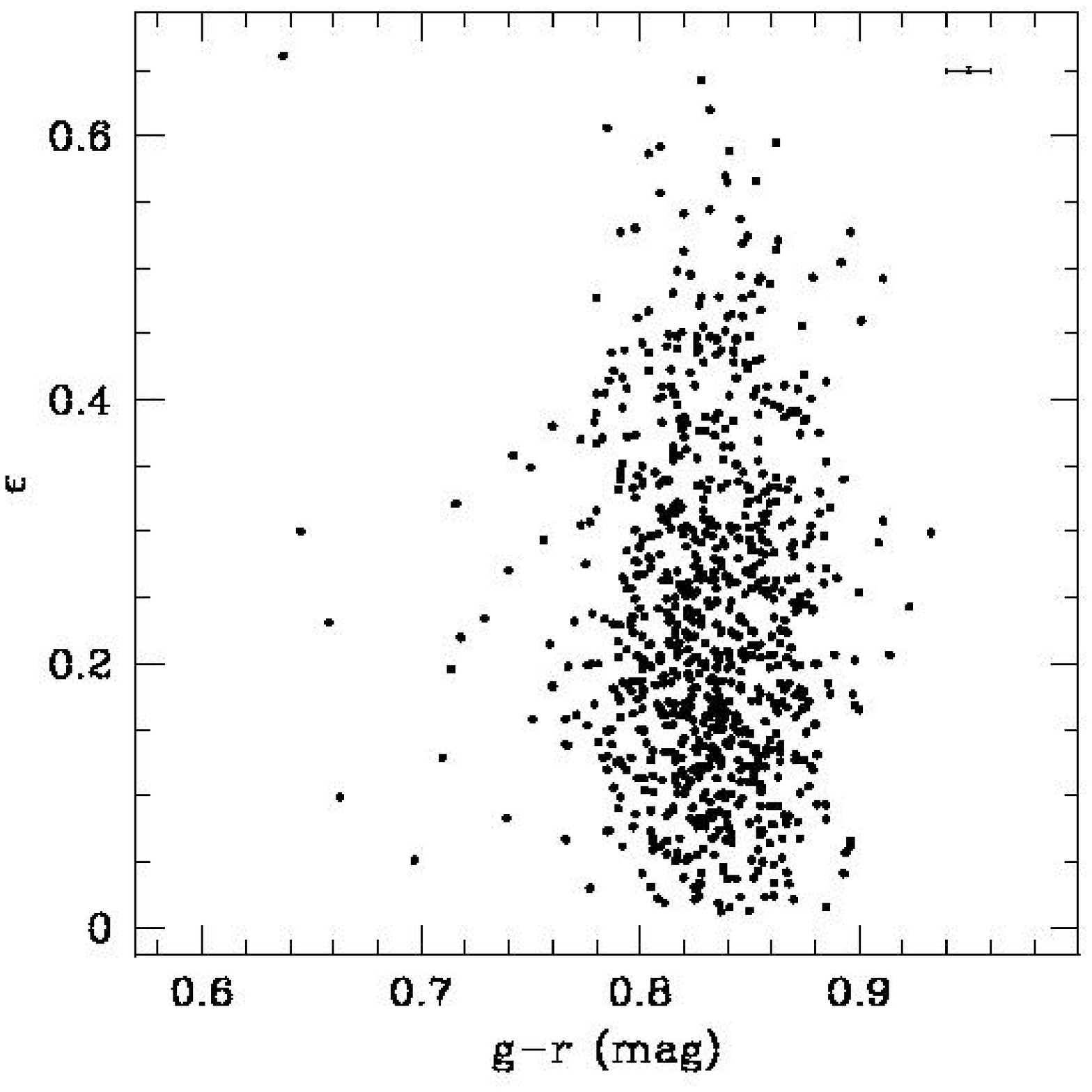}}
\caption{Mean ellipticity, $\bare$, as a function of color, $g-r$.}
\label{fig:eColor}
\end{figure*}

Fig.~\ref{fig:a3a4color} shows $a_3/a$ and $a_4/a$ versus the $g-r$ colour.
Neither shows any strong correlation. Furthermore, 
Fig.~\ref{fig:eColor} shows that the mean ellipticities of 
galaxies also have virtually no correlation with their colours.

It appears somewhat puzzling why both the mean ellipticity and $a_4/a$
are significantly correlated with $r$-band absolute magnitude while 
they do not correlate with the colour or velocity dispersion, as
if ellipticals/S0's follow the colour-magnitude relation and 
the Faber-Jackson relation ($L \propto \sigma^4$).
However, these may be explained by the large scatters in these correlations 
-- mean ellipticity versus $M_{\rm r}$ (Fig.~\ref{fig:eSigma}), $a_4/a$
versus $M_{\rm r}$  (Fig.~\ref{fig:a3a4mag}), colour-magnitude relation
(e.g., Chang et al. 2006) and the Faber-Jackson relation (e.g., Bernardi et al. 2003b).
Our sample selection criterion may also be partly responsible
for the absence of correlations of the mean ellipticity and $a_4/a$ with velocity
dispersion since our velocity dispersion is restricted to within a
factor of two (\S\ref{sec:samsel}).

Ryden et al. (2001) found that the ages of elliptical galaxies are correlated
with their ellipticities and luminosity profiles (`core' or `power-law').
We cross-correlate our sample objects with DR4 catalogue generated by 
Kauffmann et al. (2003) and find 345 objects in common. They derived
the age using a Bayes method by exploring the D$_{\rm n}$(4000), H$\delta_{\rm A}$
and a library of star formation histories generated by the stellar population
synthesis code provided by Bruzual \& Charlot (2003; See Kauffmann et al. 2003 
for details). We used here the median estimates of the $r$-band weighted mean stellar 
age in their table. However, no strong correlations 
are found between the age versus ellipticity and $a_4/a$.

As mentioned before, the s\'ersic index $n$ from surface brightness
fitting is available from Blanton et al. (2005). We 
correlated this parameter with other isophotal parameters. No correlation
was found between $n$ and $a_3/a$. However, there appears to be a
correlation between $n$ and $a_4/a$: the Spearman-Rank Order
correlation coefficient is $-0.104$ and the probability that no
correlation exists between these two parameters is only 0.24 per cent. Galaxies
with large s\'ersic index $n$ tend to have smaller deviations from perfect
ellipses. There is also a weak trend that more flattened (i.e. larger
ellipticity) galaxies have smaller values of $n$, but the scatter is quite large.

\subsection{Isophotal twists}

It is well known that isophotes in elliptical galaxies are not
concentric ellipses as a function of radius. In this
subsection, we concentrate on the changes of isophotes from one to
one and a half Petrosian half-light radii ($\rp \rightarrow 1.5\rp$).

\begin{figure*}
\centerline{\includegraphics[width=0.6\textwidth]{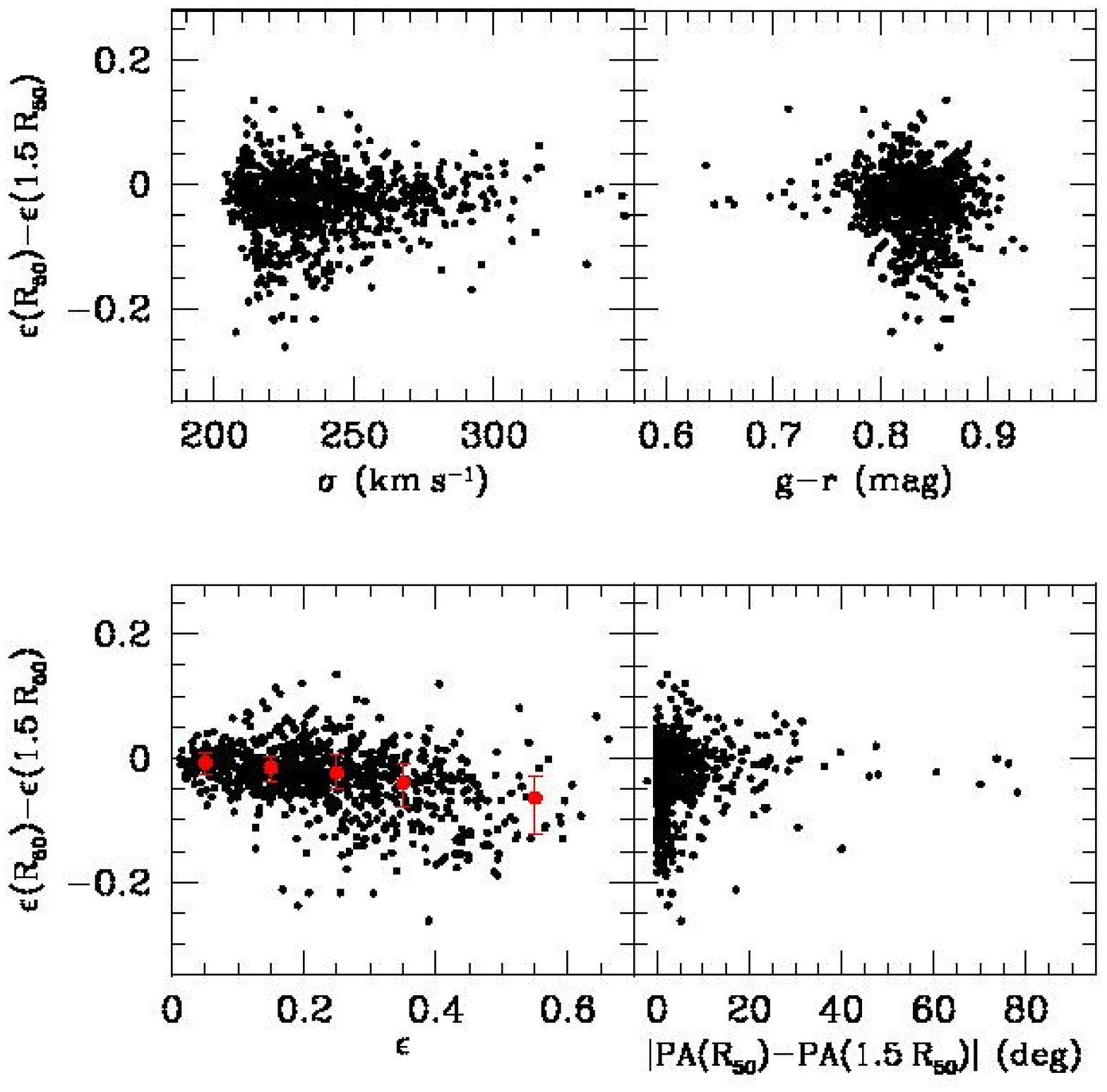}}
\caption{The difference in the ellipticity between one and one and a half
Petrosian half-light radii ($\rp$ and $1.5\,\rp$) for our sample galaxies vs. velocity dispersion,
  colour, mean ellipticity, and change in the position
  angle from $\rp$ to $1.5\,\rp$. The data points with error bars are the median
and lower (25 per cent) and upper (75 per cent) quartiles for galaxies in bins of width 0.1
in ellipticity except
the last bin with a width 0f 0.3 to include all the remaining objects.
}
\label{fig:eTwist}
\end{figure*}

Fig.~\ref{fig:eTwist} shows the change of ellipticity as a function
of the velocity dispersion, colour, mean ellipticity, and the difference
in the position angle between $\rp$ and $1.5\,\rp$. There is no strong correlation between
the change and either velocity dispersion, colour or the change in position 
angle.
There is however a correlation between the mean ellipticity $\bare$
and the ellipticity change, with Spearman-Rank Order correlation coefficient
$r_s=-0.373$ at a significance level of $>99.9$ per cent.
Elliptical galaxies with higher ellipticities tend to have
a more negative value of $\delta \epsilon = \epsilon(\rp)-\epsilon(1.5\,\rp)$. 
This implies that the isophotes in the most elliptical galaxies become more 
flattened as one goes to larger radii. 

\begin{figure*}
\centerline{\includegraphics[width=0.6\textwidth]{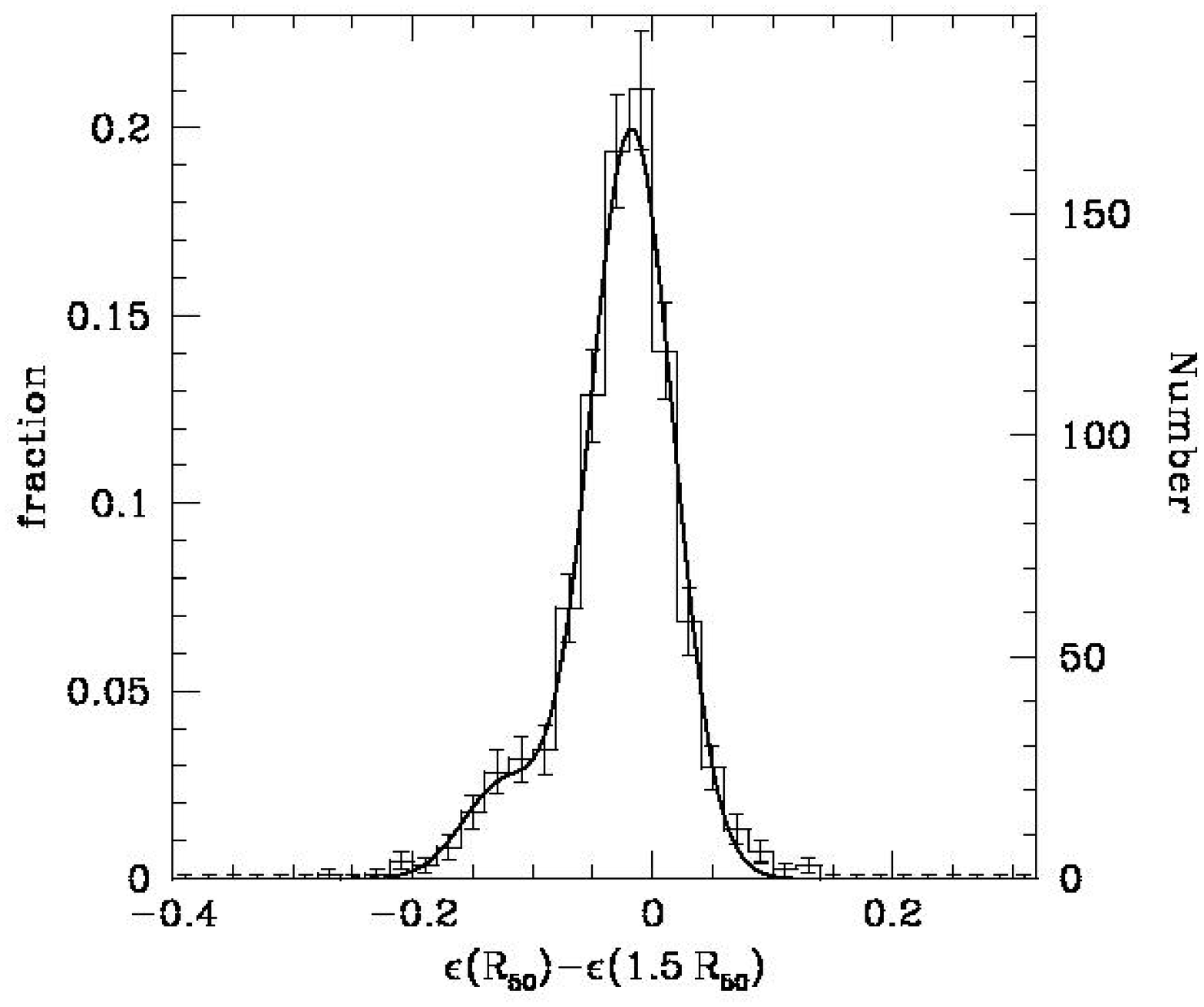}}
\caption{Histogram of the difference in the ellipticity between one and one
and a half
  Petrosian half-light radii for our sample galaxies. The solid line is a double Gaussian
  fit to the data.
}
\label{fig:deHist}
\end{figure*}

Fig.~\ref{fig:deHist} shows the histogram of the ellipticity change, $\delta\epsilon$.
The mean change in ellipticity is about $-0.03$ and the scatter is about 0.051.
The differential probability distribution can be well-fitted by the sum of two Gaussians:
\beq
{dP \over d\delta\epsilon} = 
 {0.017\over \sqrt{2 \pi} \sigma_1} {\rm
   e}^{-{(\delta\epsilon-\delta\epsilon_1)^2 \over 2 \sigma_1^2} } +
 {0.002 \over \sqrt{2 \pi} \sigma_2} {\rm e}^{-{(\delta\epsilon-\delta\epsilon_2)^2 \over 2 \sigma_2^2} }
\eeq
where $\delta\epsilon_1=-0.018$, $\sigma_1=0.035$,
$\delta\epsilon_2=-0.124$, and $\sigma_2=0.034$. This fit is shown
as the solid curve. The second peak is due to the
galaxies with high ellipticities (see the bottom left panel of Fig.~\ref{fig:eTwist}).

\begin{figure*}
\centerline{\includegraphics[width=0.6\textwidth]{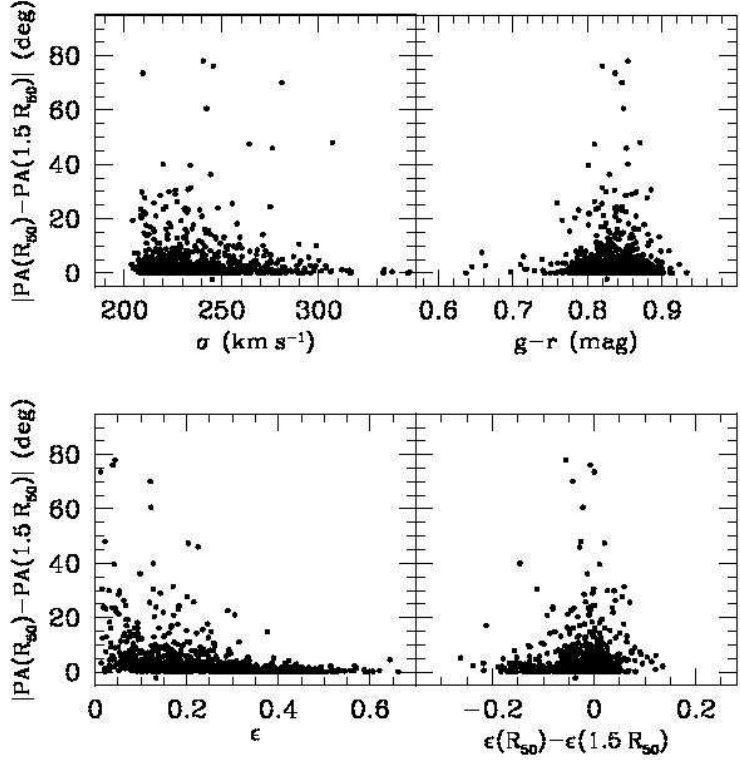}}
\caption{The difference in the position angle between one and one and a half
  Petrosian half-light radii for our sample galaxies vs. velocity dispersion,
  colour, mean ellipticity, and change in the
  ellipticity between $\rp$ and $1.5\,\rp$.
}
\label{fig:PAtwist}
\end{figure*}

\begin{figure*}
\centerline{\includegraphics[width=0.6\textwidth]{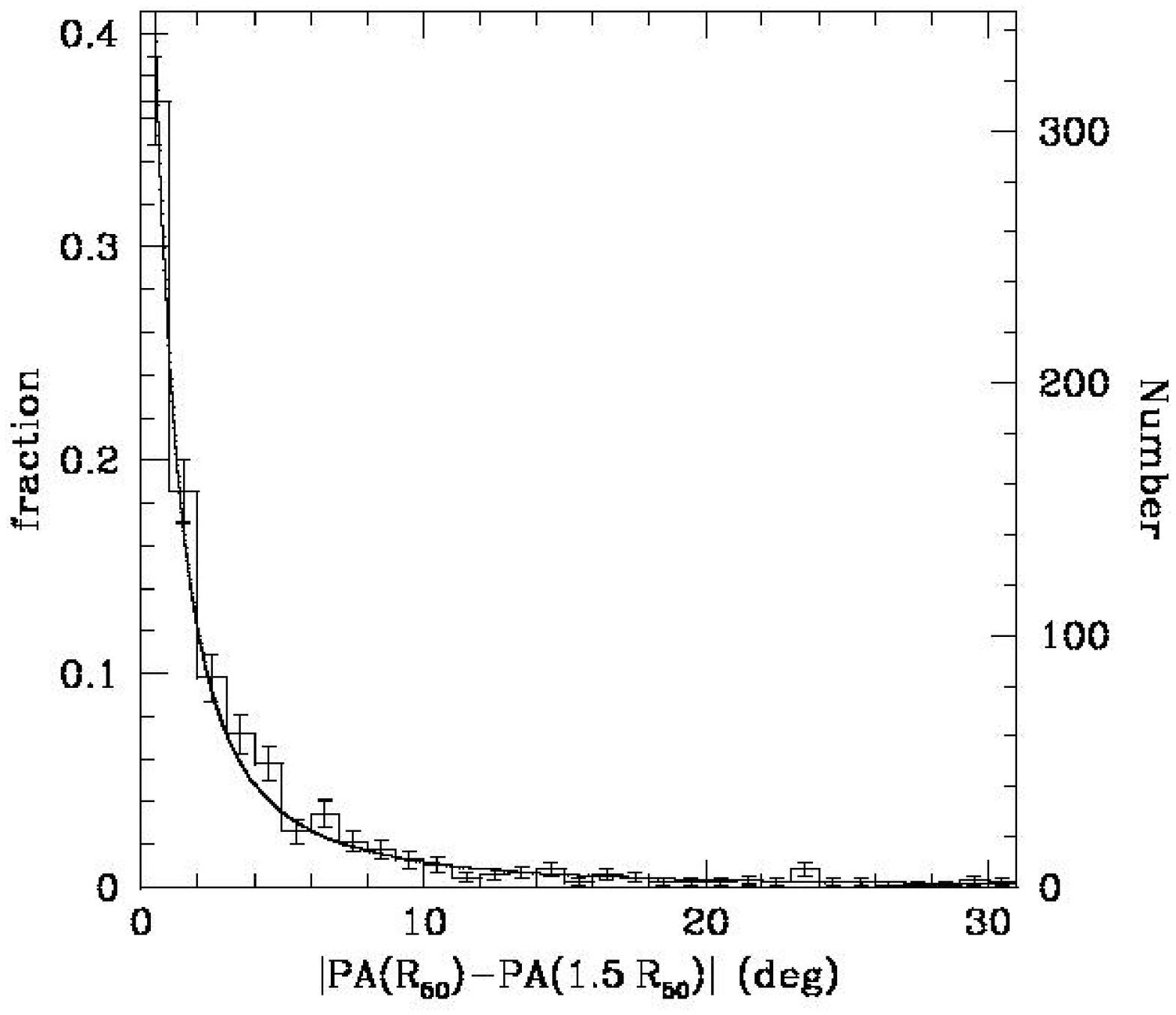}}
\caption{Histogram of the difference in position angle between one and one and 
a half Petrosian half-light radii for our sample galaxies. The solid line is a 
power-law fit to the data; the power-law index is $-$1.81.}
\label{fig:dPAhist}
\end{figure*}

Fig.~\ref{fig:PAtwist} shows the change of position angle as a function
of the velocity dispersion, colour, mean ellipticity, and the difference
in ellipticity between $\rp$ and 1.5$\rp$. In this case, there is no strong 
correlation with all these quantities. Fig.~\ref{fig:dPAhist} shows the 
histogram of the changes of the position angles. The differential probability 
distribution can be well fitted by a power-law
\beq
{dP \over \dPA} = (\beta-1) \left(1+{\dPA \over {\dPA}_0}\right)^{-\beta},
\eeq
where $\delta{\rm PA}_0=1.07^\circ$ and $\beta=1.81$. The median 
value of $\delta{\rm PA}$ is $1.61^\circ$ . In contrast,
the mean change in position angle is much larger, about $4.12^\circ$
and the standard deviation is about $8.39^\circ$. The large standard deviation 
is due to an extended tail in $\delta{\rm PA}$.

\begin{figure*}
\centerline{\includegraphics[width=0.6\textwidth]{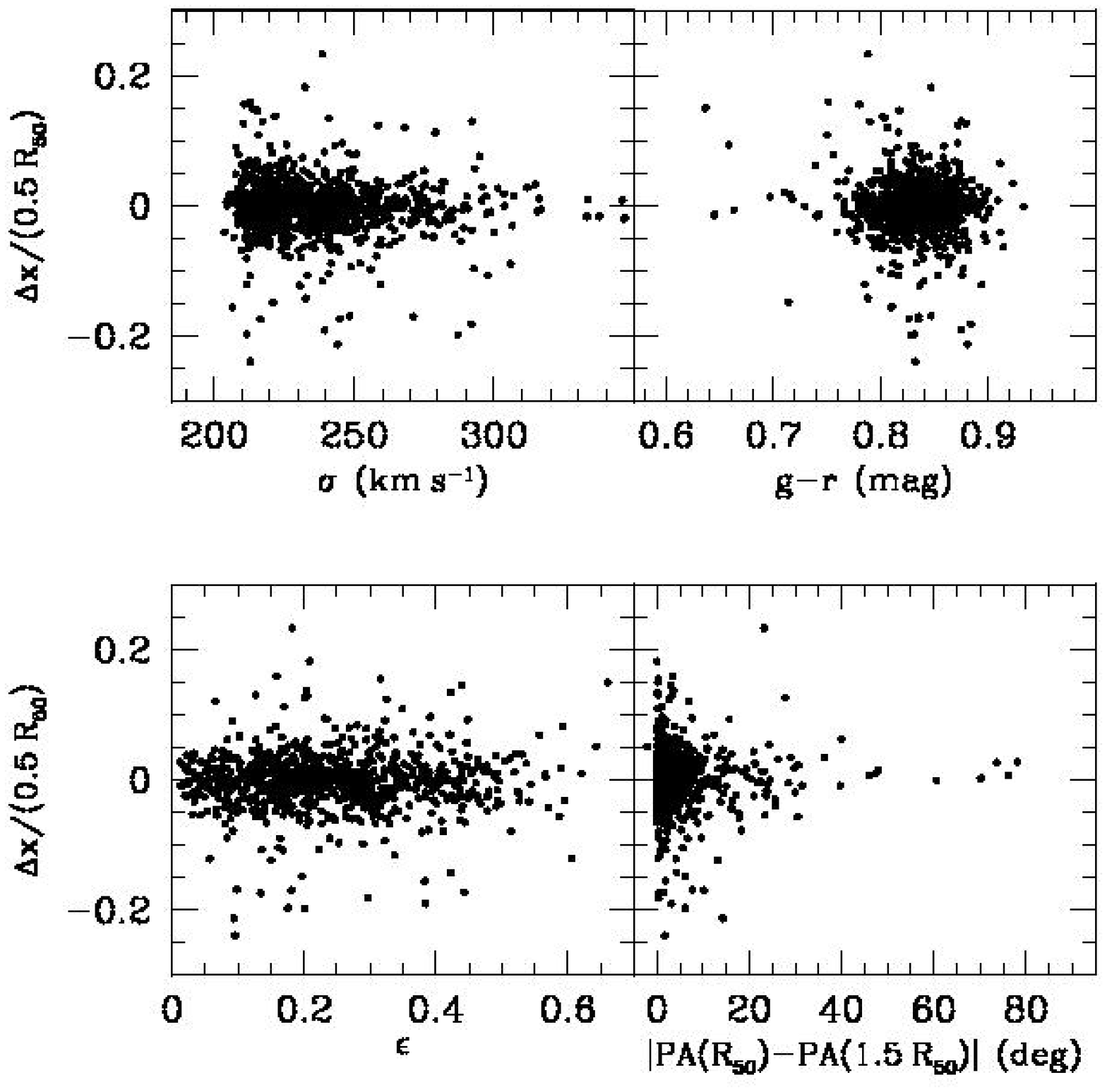}}
\caption{Centroid shift (in units of half a Petrosian half-light radius) from
one to one and a half Petrosian half-light radii
in the major-axis direction vs. velocity dispersion,
  colour, mean ellipticity, and change in the position
  angle between $\rp$ and $1.5\,\rp$.
}
\label{fig:dCenterx}
\end{figure*}

\begin{figure*}
\centerline{\includegraphics[width=0.6\textwidth]{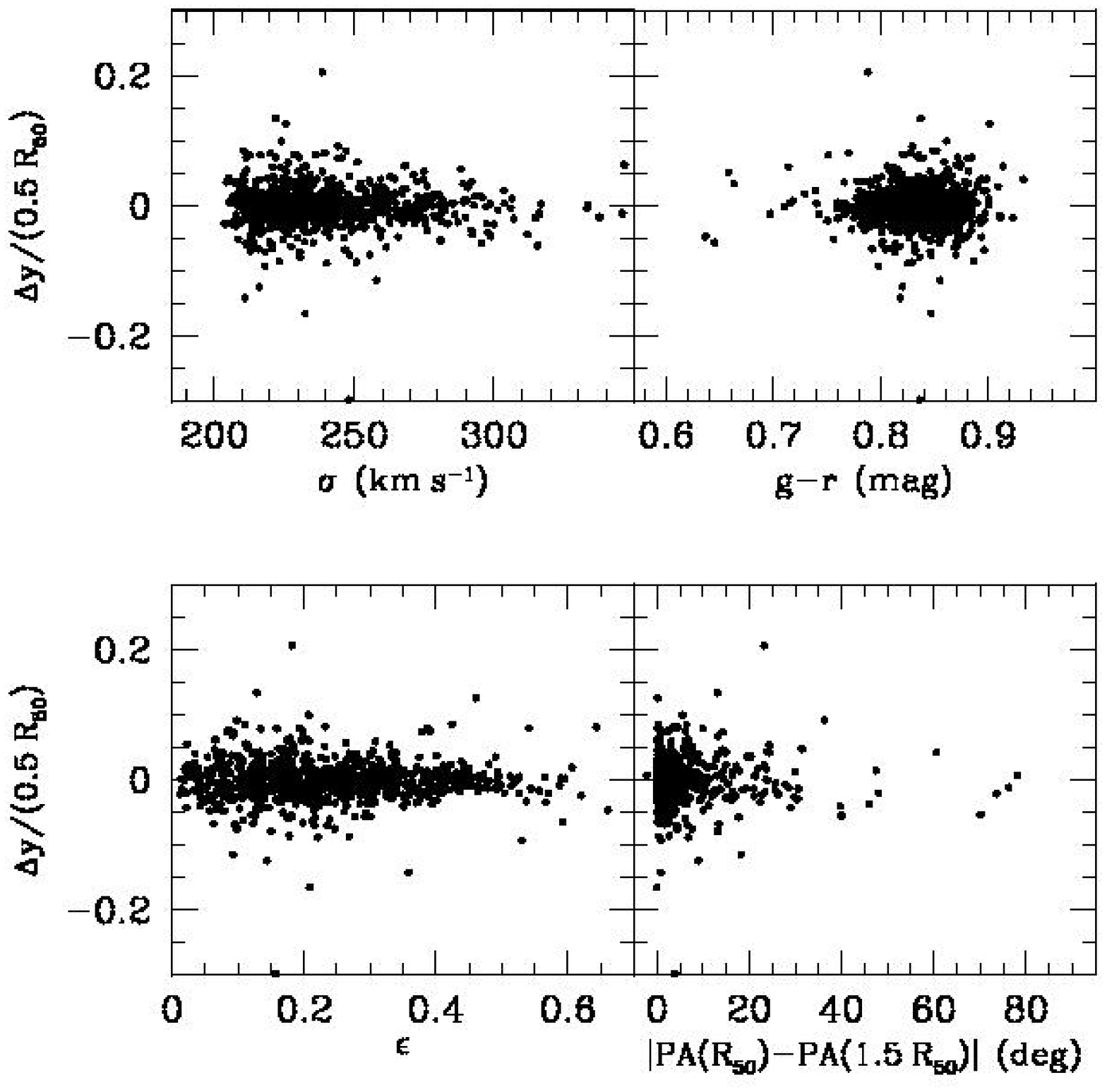}}
\caption{Centroid shift (in units of half a Petrosian half-light radius) from
one to one and a half Petrosian half-light radii
in the minor-axis direction vs. velocity dispersion,
  colour, mean ellipticity, and change in the position
  angle between $\rp$ and $1.5\,\rp$.}
\label{fig:dCentery}
\end{figure*}

\begin{figure*}
\centerline{\includegraphics[width=0.6\textwidth]{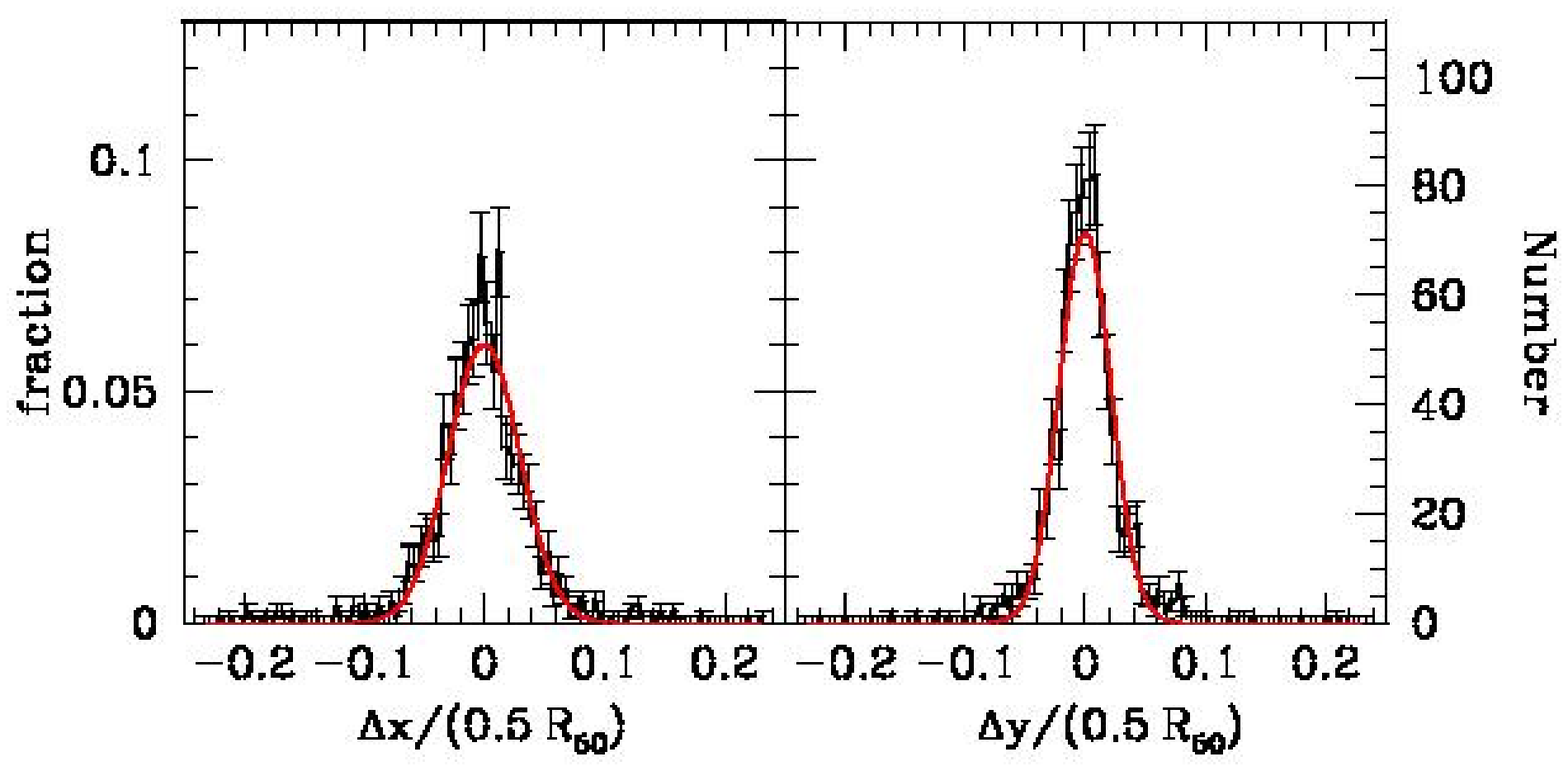}}
\caption{Histograms of the centroid shifts from
one to one and a half Petrosian half-light radii
in the major (left) and minor (right)
axis directions in units of half a Petrosian half-light radius. The solid 
lines are the best-fitting Gaussians.}
\label{fig:dCenterHist}
\end{figure*}

Figs.~\ref{fig:dCenterx} and \ref{fig:dCentery} show the changes of the 
centres of ellipses along the major (x) and minor (y) axes as a function of velocity dispersion, colour, 
mean ellipticity and the difference in position angle between $\rp$ and $1.5\,\rp$.
There is no significant correlation among these quantities. The
histograms of the changes of the centres 
of ellipses are shown in Fig.~\ref{fig:dCenterHist}. The differential
probability distribution can be fitted by a standard Gaussian with zero
means and
$\sigma_{{\delta \rm x}/(0.5\,\rp)}=0.030$ and $\sigma_{{\delta \rm y}/(0.5\,\rp)}=0.021$
respectively. The centroid shifts are hence small, only a few percent of the Petrosian half-light radius.

\subsection{Radio properties of elliptical galaxies}

We search the radio counterparts of our early-type galaxies using
the FIRST (Becker, White \& Helfand 1995) and NVSS (Condon et al. 1998) catalogues. 
The NVSS and FIRST surveys provide an angular resolution (beam size) of
45$\arcsec$ and 5$\arcsec$ respectively. However, the centroids can be determined much
more accurately and the accuracy scales roughly as $1/\sqrt{{\rm signal-to-noise\ ratio}}$.
We adopt a conservative cross-correlation radius between the radio catalogue and our sample objects.
We found 162 counterparts for FIRST within a 5$\arcsec$ cross-correlation radius and 130 for
NVSS galaxies within a 20$\arcsec$ cross-correlation radius.

\begin{figure*}
\centerline{\includegraphics[width=0.6\textwidth]{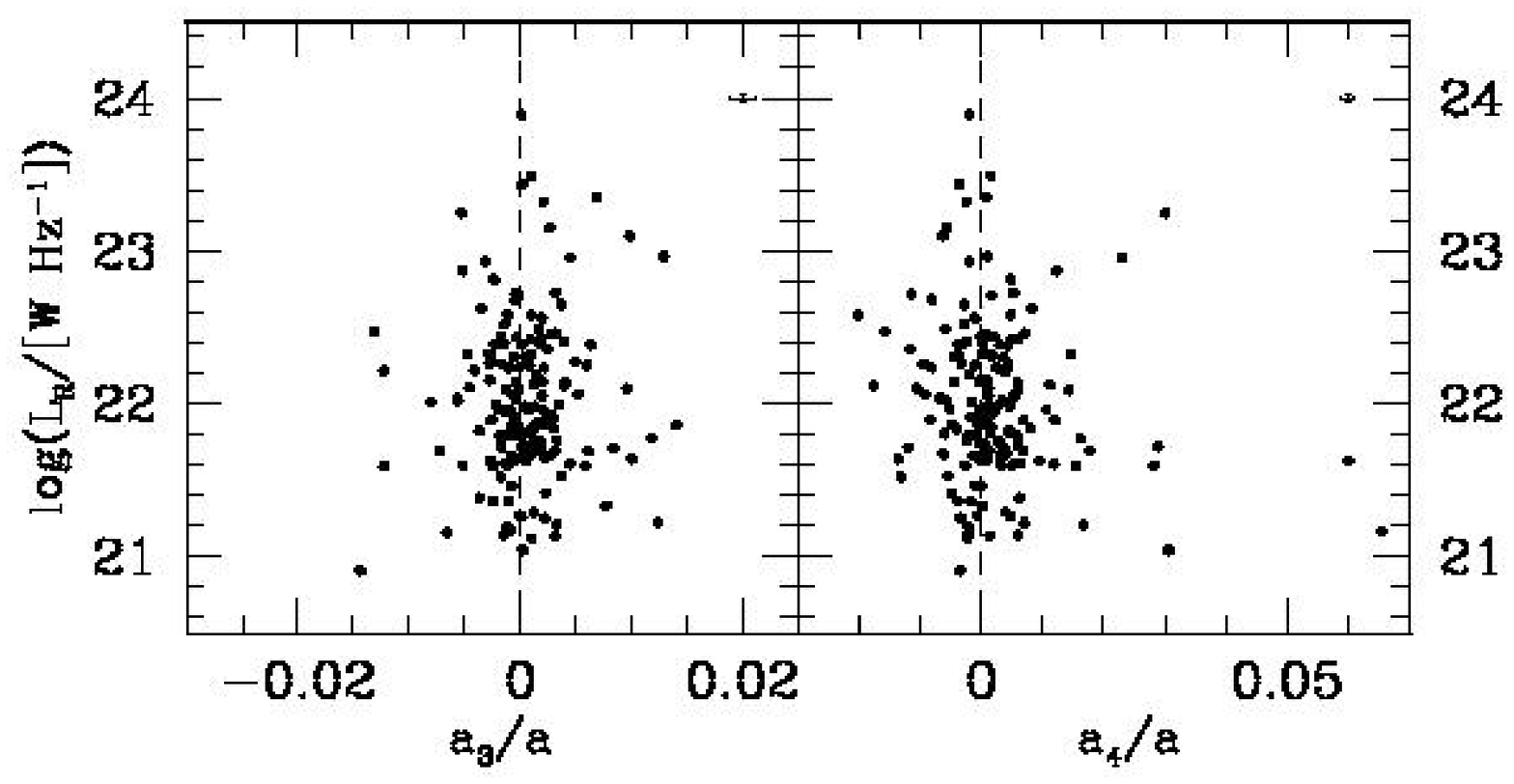}}
\caption{Radio power vs. $a_3/a$ and $a_4/a$ parameters for elliptical
and S0 galaxies identified from a cross-correlation between the
SDSS and the FIRST radio catalog.}
\label{fig:first}
\end{figure*}

As the position accuracy for FIRST is higher, here we only present the
results for the objects cross-correlated from FIRST and SDSS.
Fig.~\ref{fig:first} shows the 1.4GHz FIRST peak luminosity
as a function of the $a_3/a$ parameter and $a_4/a$ parameter.
There is no trend between
the radio power and the $a_3/a$ parameter. For the $a_4/a$ parameter
(shown in the right panel), there
appears to be a lack of galaxies at the top right corner. (The trend is
similar for the 1.4GHz
FIRST integrated luminosity versus $a_4/a$.) Thus
we confirm the trend found by Bender et al. (1989) that discy
ellipticals (with $a_4>0$) are weaker radio emitters than their
boxy counterparts, although there are much less fainter radio emitters in our
sample compared to Bender et al. (1989) because of the larger redshift
in our sample relative to theirs.

\subsection{Correlations with environment}

Di Tullio (1979) found that the ellipticity of isolated ellipticals
on average decreases as a function of radius, while for ellipticals in clusters 
and groups, the ellipticity can either increases, decreases or peaks. This 
conclusion was based
on photographic plates for 75 ellipticals in clusters, groups, pairs
and isolated galaxies. More recently, using the SDSS data, Kuehn \& Ryden (2005)
concluded that as the local density of galaxies increases, galaxies with de 
Vaucouleurs profiles become rounder in the magnitude range of
$-22<M_{\rm r}<-18$.
On the other hand, for the most luminous galaxies ($M_{\rm r}<-22$),
galaxies become rounder in the inner regions, but exhibit no change in
their outer regions as the local density of galaxies increases. 
Shioya \& Taniguchi (1993) used 96 galaxies in Bender et 
al. (1989) to study the isophotal shape-environment relation and found that
boxy-type galaxies favor the dense environments while discy-type galaxies prefer
the field environments.

In this paper, following Kuehn \& Ryden (2005), we use a measure of the local
density as the number of galaxies within a cylinder in the redshift space. 
The depth of the cylinder (centred on a target) is $12h^{-1}\mpc$, and we vary
the radius of the cylinder from $0.3h^{-1}\mpc$ to $0.5h^{-1}\mpc$,
where $h=H_0/(100 \kms\mpc)=0.7$.  
According to Kuehn \& Ryden (2005), if a target galaxy is within one cylinder
width of the SDSS spectroscopic survey border, it is removed from the analyses.
The radii of the cylinders we adopted above are smaller than those used by Kuehn \& Ryden (2005)
because if we adopt their larger values (e.g., $2h^{-1}\mpc$), too few galaxies would be left
because our sample galaxies are at lower redshift and so the
boundary effects discussed above would exclude too many galaxies.
Following Kuehn \& Ryden (2005),
we define a volume limited sample which is selected from SDSS DR4 catalogue
by constraining the spectral redshift smaller than 0.055. For these objects,
we also adopt redshifts corrected for the Local Group infall provided by Blanton et al. (2005).
As the maximum redshift of our sample galaxies
is 0.05 and the limiting apparent magnitude of the spectroscopic survey is
$m_{\rm r}=17.77$, the corresponding absolute magnitude is $M_{\rm r}=-19$.
We use the number counts of galaxies with $M_{\rm r} < -19$ within
a cylinder of the target galaxy as an indicator of the local density.
This is only an approximate measure of the local environment, due to the
effects of fiber collisions (see Kuehn \& Ryden 2005 for detailed
discussions) and redshift distortions. But it is sufficient for our purpose.
For brevity, we term galaxies without neighbours in the cylinder as
`isolated' galaxies.

\begin{figure*}
\centerline{\includegraphics[width=0.6\textwidth]{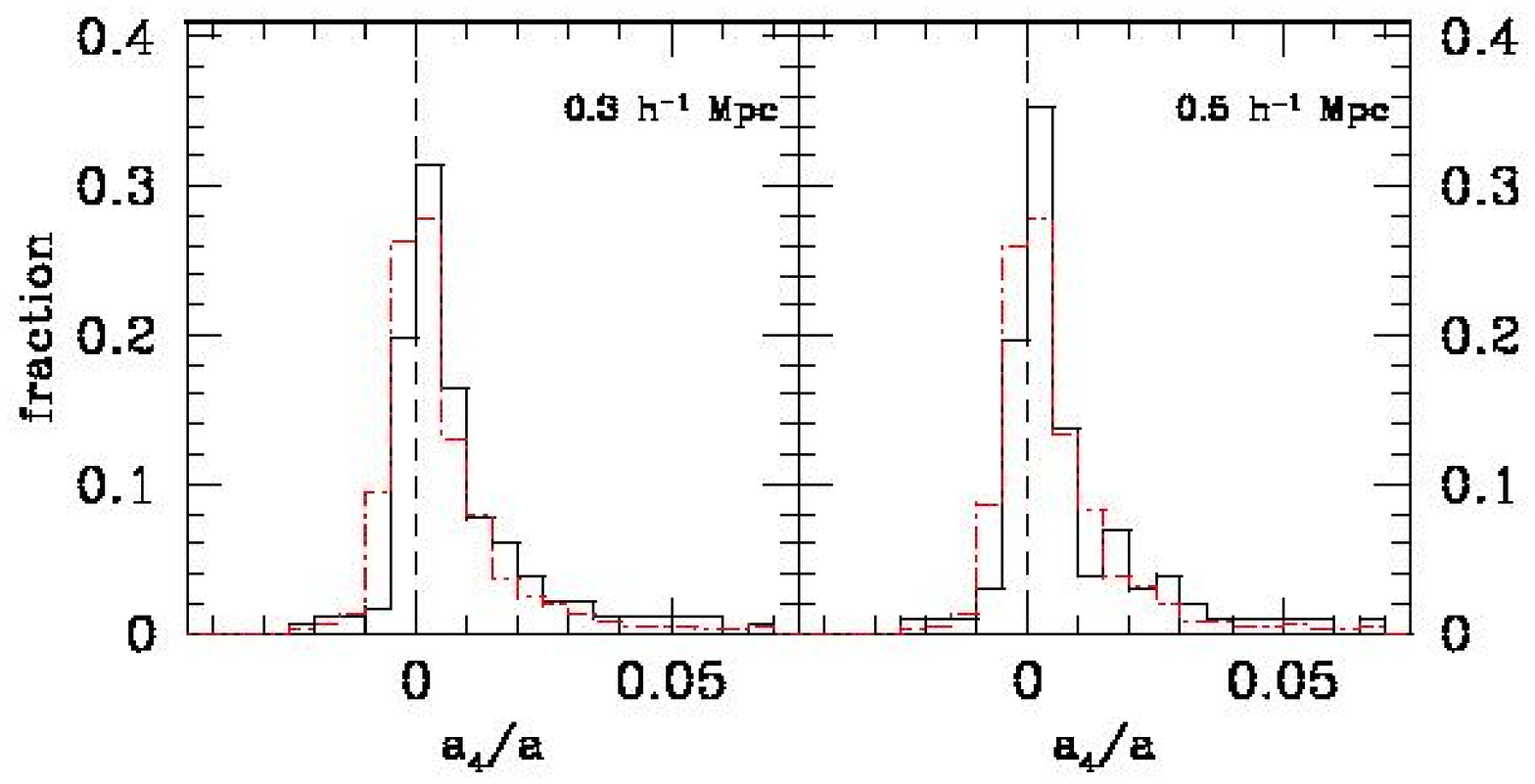}}
\caption{Histograms of $a_4/a$ for isolated galaxies (solid curve)
and galaxies with neighbour (dot-dashed curve). Isolated galaxies are those
with zero neighbour within a cylinder of depth $12h^{-1}\mpc$ and radius
$0.3h^{-1}\mpc$ (left) and $0.5h^{-1}\mpc$ (right). For clarity, Poisson errors
are not plotted.}
\label{b4mean.envhis.eps}
\end{figure*}

Fig.~\ref{b4mean.envhis.eps} shows the $a_4/a$ distribution
for `isolated' galaxies (solid curve) and galaxies with neighbour (dot-dashed
curve). The left and right panels are for a cylinder 
with radius of $0.3h^{-1}\mpc$ and $0.5h^{-1}\mpc$  respectively.
We can see that the fraction of discy galaxies is larger in isolated 
environments. Quantitatively, 
for cylinders with radii of $0.3h^{-1}\mpc$ (left panel) and $0.5h^{-1}\mpc$ 
(right panel), the ratios of the number of discy to
boxy E/S0s in `isolated' environments are 3.14 (138/44) and 2.92 (76/26) 
respectively.
In contrast, for galaxies with neighbours, the corresponding ratios are
1.63 (381/234) and 1.72 (386/225). Clearly discy E/S0s are
more abundant in isolated environments, in
agreement with the conclusion of Shioya \& Taniguchi (1993). The 
Kolmogorov--Smirnov test indicates a
probability of 99.9 (90.2) per cent that the two samples are 
different for cylinders with radii of $0.3h^{-1}\mpc$ ($0.5h^{-1}\mpc$).

\begin{figure*}
\centerline{\includegraphics[width=0.6\textwidth]{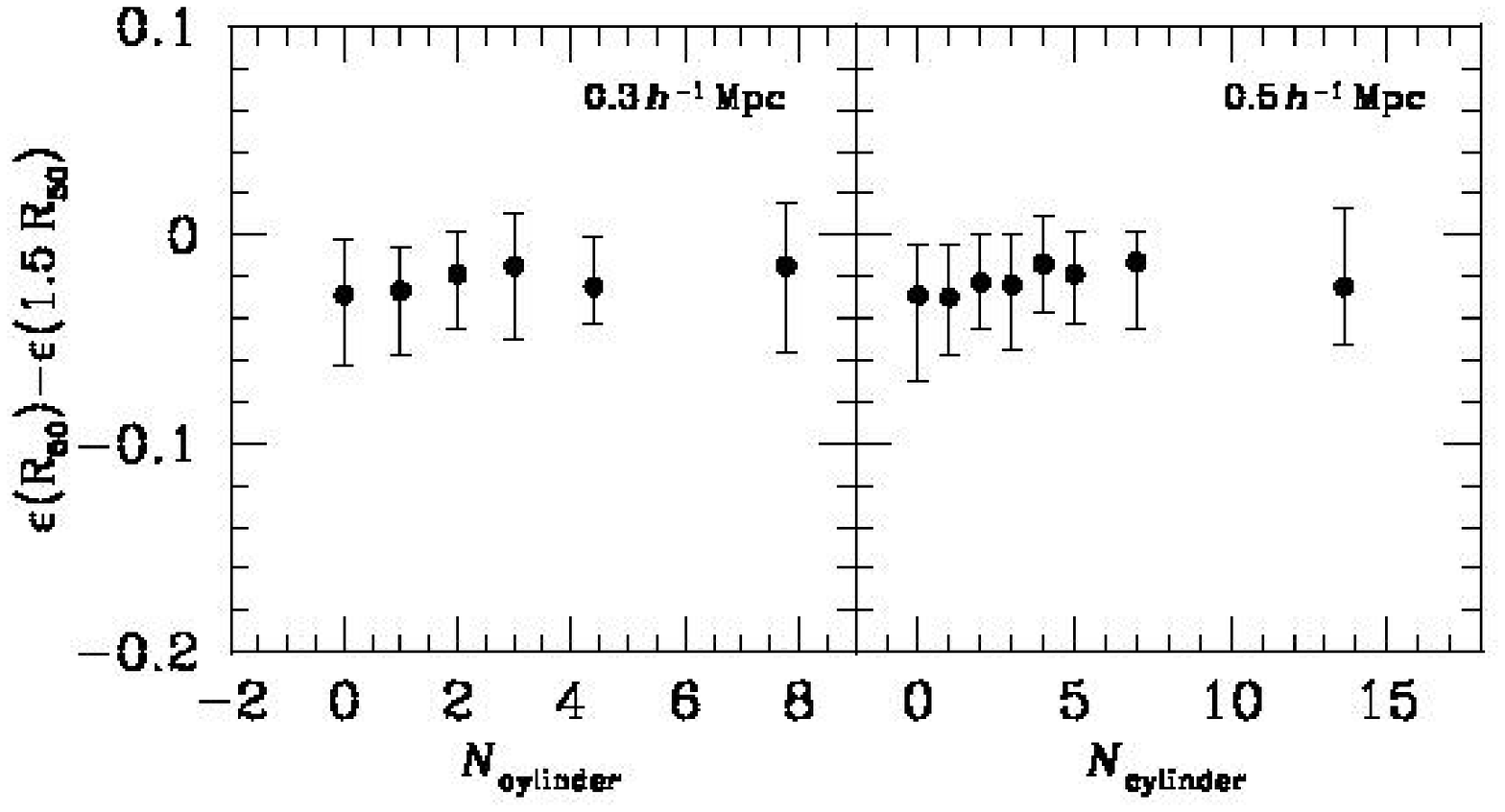}}
\caption{The change of ellipticity between one and one and a half Petrosian 
half-light radii as a function of the local density, as measured by the number 
of galaxies, $N_{\rm cylinder}$, within a cylinder of depth  $12h^{-1}\mpc$ and 
radius $0.3h^{-1}\mpc$ (left) and $0.5h^{-1}\mpc$ (right). The data points
with error bars are the median, lower (25 per cent) and upper (75 per cent) 
quartiles for bins of $N_{\rm cylinder}$.}
\label{elliptwist.env.eps}
\end{figure*}

\begin{figure*}
\centerline{\includegraphics[width=0.6\textwidth]{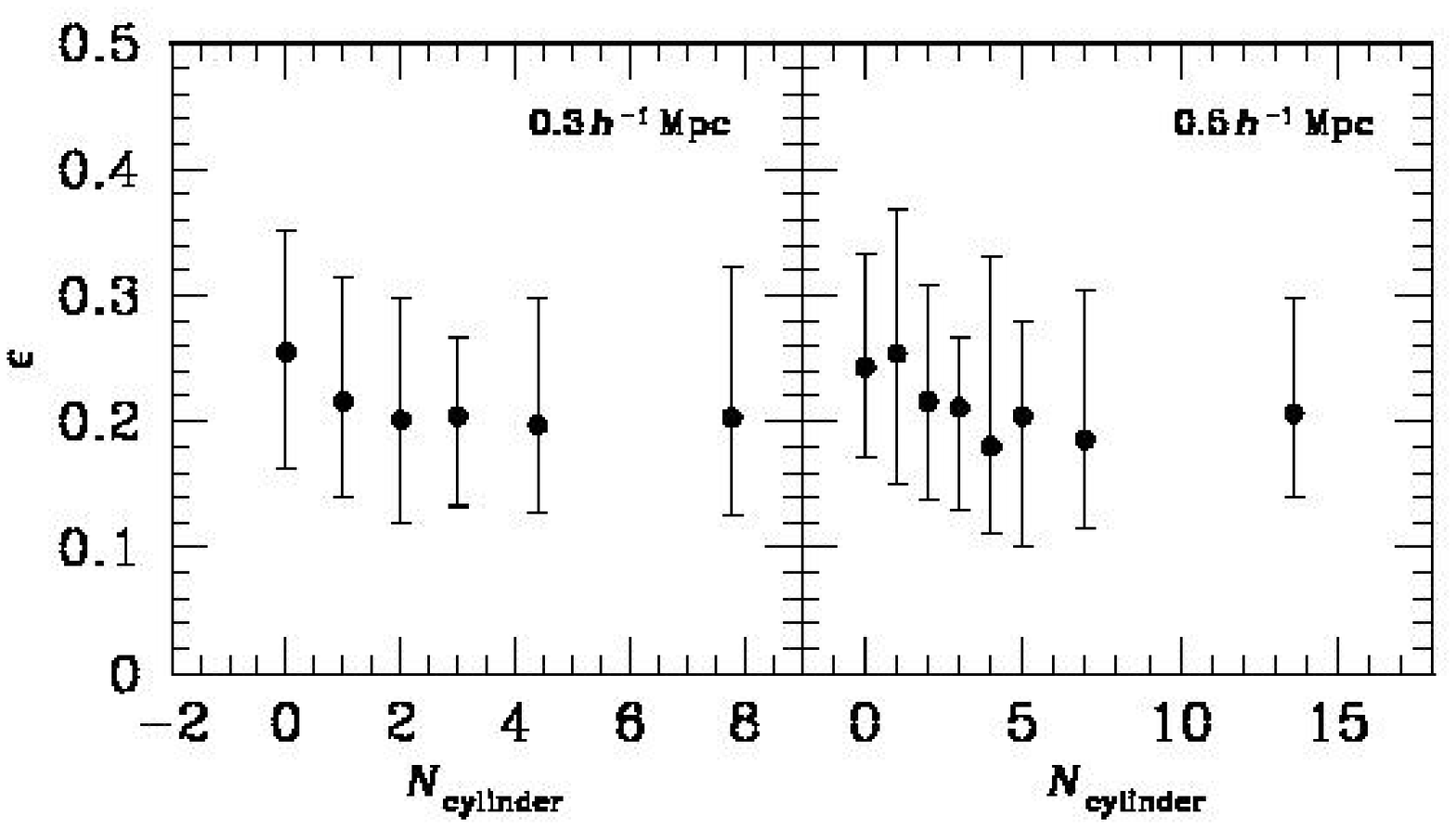}}
\caption{Ellipticity versus local density, 
as measured by the number of
galaxies within a cylinder of depth  $12h^{-1}\mpc$ and radius
$0.3h^{-1}\mpc$ (left) and $0.5h^{-1}\mpc$ (right). The data points
with error bars are the median, lower (25 per cent) and upper (75 per cent) 
quartiles for bins of $N_{\rm cylinder}$.}
\label{ellipmean.env.eps}
\end{figure*}

The change of ellipticity as a function of the local density is shown in
Fig.~\ref{elliptwist.env.eps}.
There seems no correlation between the change of ellipticity and the local 
density. This result is not consistent with Di Tullio (1979). 
His sample based on photographic plates may be too small to be statistically convincing.
Reda et al. (2004) presented the ellipticity profiles for 14 isolated 
early-type galaxies (their fig.~4); their results also do not support
the conclusion of Di Tullio (1979).
Fig.~\ref{ellipmean.env.eps} shows the mean ellipticity as a function of
the local density, a plot similar to figs.~6 and 7 in Kuehn \& Ryden (2005). 
Qualitatively, our results are consistent with Kuehn \& Ryden (2005).

\begin{figure*}
\centerline{\includegraphics[width=0.6\textwidth]{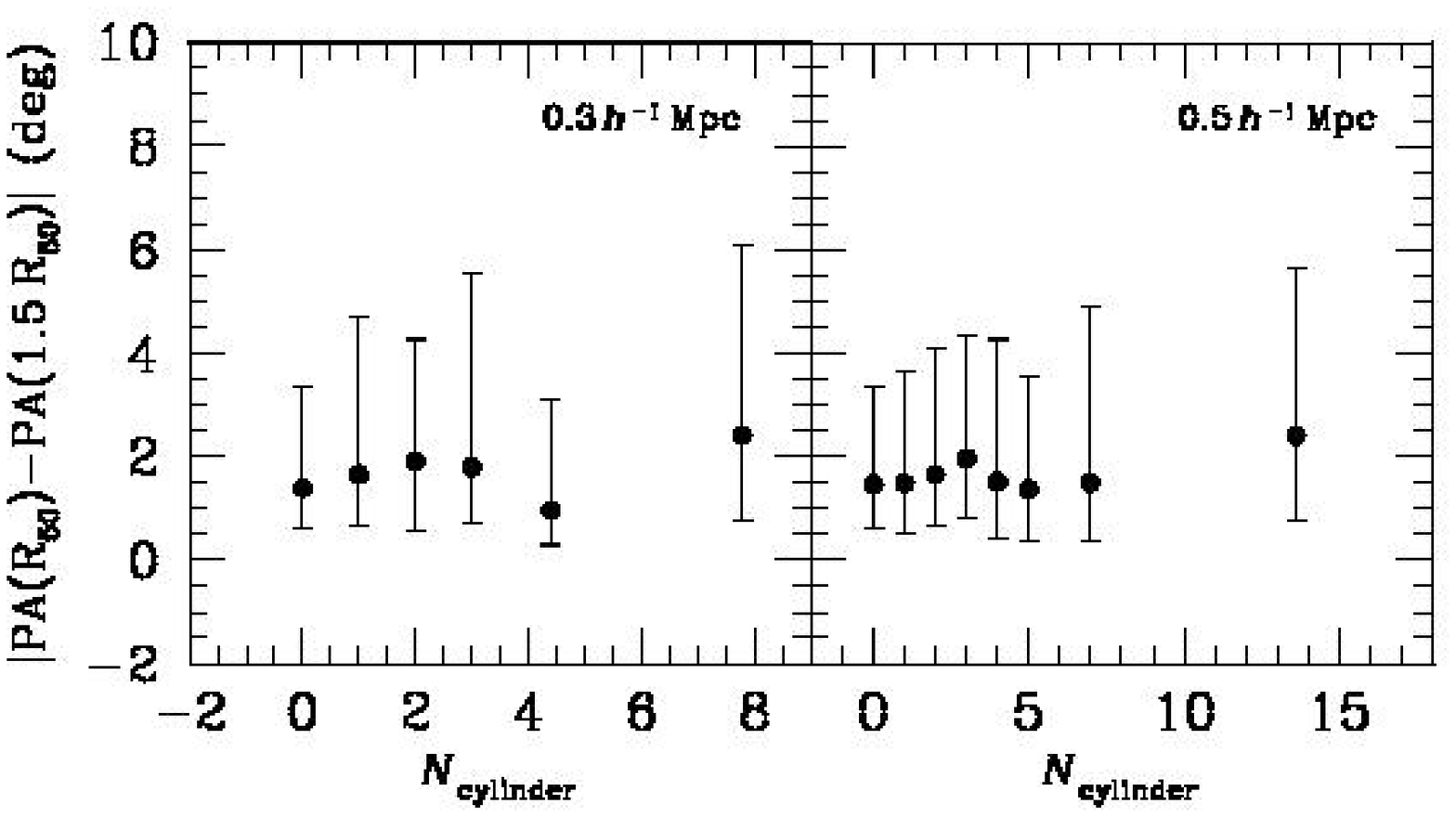}}
\caption{The twist of position angle from one to one and a half
Petrosian half-light radii as a function of local density,
as measured by the number of
galaxies within a cylinder of depth  $12h^{-1}\mpc$ and radius
$0.3h^{-1}\mpc$ (left) and $0.5h^{-1}\mpc$ (right). The data points
with error bars are the median, lower (25 per cent) and upper (75 per cent) 
quartiles for bins of $N_{\rm cylinder}$.}
\label{patwistacu.env.eps}
\end{figure*}

 As isophotal twists may be caused by external perturbations, we
examined the isophotal twist - local density relation in 
Fig.~\ref{patwistacu.env.eps}. However, there is no correlation found.
Alternatively, we also examined the ratio of the number of E/S0s with large PA 
changes between $\rp$ and 1.5$\rp$ to that of E/S0s with small PA changes
for isolated galaxies and galaxies with neighbours.
The results show that the fraction of E/S0s with large PA changes is slightly
larger for galaxies with neighbours than that for isolated galaxies.

\section{Conclusions and Discussions}

In this paper, we studied the shapes of isophotes for a sample of
847 nearby E/S0 galaxies selected from the SDSS DR4
catalogue. We derived the distributions
of the intensity-weighted ellipticity and Fourier coefficients and also
studied the changes in the isophote shapes between one and one and a half Petrosian 
half-light radii. For many of these distributions, we provide empirical fitting
formulas of the data, which may be useful for comparisons with
theoretical studies. We further examined many possible correlations
between the isophote shapes and physical properties of E/S0 galaxies.
Our main conclusions are the following:
\begin{enumerate}
\item
The mean ellipticity of galaxies is about 0.23 with a scatter of 0.13. 
This is somewhat narrower than the distribution found by Lambas et al. (1992).
The intensity-weighted mean ellipticity shows no strong correlation 
with either the velocity dispersion or the $g-r$ colour. However,
it is correlated with the $r-$band absolute magnitude and dynamical mass.
These correlations indicate that the more luminous (massive) E/S0s are
rounder than the fainter (less massive) ones. 
\item
The most significant deviation of the isophotes from
ellipses is the $a_4/a$ parameter determined
from Fourier series. This parameter shows no significant correlation
with colour and velocity dispersion. It is, however, correlated
with the $r-$band absolute magnitude, in agreement with
Bender et al. (1989). As the luminosity decreases, the number of
discy E/S0's increases relative to boxy ones.
The $a_4/a$ parameter is also strongly correlated
with the mean ellipticity (Fig. \ref{fig:a4e}).
E/S0's with higher ellipticities tend to show stronger deviations in isophotes 
from perfect ellipses. They can become either more discy or more boxy. This suggests
that the isophotes are likely a
function of viewing angle -- edge-on ellipticals will show stronger
deviations from ellipses.
\item
Our studies on the relation between isophote shapes and environments show
that discy E/S0s favor field environments while boxy E/S0s prefer
dense environments, in agreement with Shioya \& Taniguchi (1993).
The changes of ellipticity and position angle with radius, however,
have no significant correlations with the environments.
\item	
We confirm the results of Bender et al. (1989) that 
discy ellipticals tend to lack powerful radio emitters
(Fig.~\ref{fig:first}), although the trend is not dramatic.
\item
The changes in ellipticity and position angle between one and one and a half Petrosian 
half-light radii show no correlations with galaxy properties,
except that the change in ellipticity is correlated
with the mean ellipticity (see Fig.~\ref{fig:eTwist}). E/S0 galaxies with 
larger ellipticities become more flattened
at large radii. The median change in the ellipticity and
position angle is about $-$0.023 and $1.61^\circ$ respectively. The
change in the position angle has a significant tail. 
\end{enumerate}

Gravitational lensing can be used to probe the shape of total matter
distribution in intermediate redshift galaxies. 
Yoo et al. (2006) derived the limits for four lensing systems, 
SDSS\,J0924+0219, HE\,0435-1223, B\,1938+666, and PG\,1115+080. The upper limits
on $a_3/a$ are 0.023, 0.019, 0.037 and 0.035, and 
$a_4/a$ are 0.034, 0.041, 0.051, 0.064 respectively for these
galaxies. Our study can only constrain the shape of stellar matter distribution. Nevertheless,
Fig.~\ref{fig:a3a4Hist} shows that the vast majority of our elliptical galaxies
satisfy these constraints as long as the dark matter distribution
does not show more significant deviations from the elliptical
distributions. It is interesting to note that in Fig.~\ref{fig:eHist} 
the ellipticity distribution (dashed histogram) of the lens galaxies
(selected largely by mass) are consistent with our luminosity-selected
sample. The number of lenses in the Koopmans et al. study is only
15, so is still somewhat small  to draw stronger conclusions.
Another effect of gravitational lensing is that deviations from perfect
ellipses can cause uncommon lensing configurations, such as
sextuplet and octuplet imaging. The solid curves in
Fig.~\ref{fig:a4e} show the critical lines for forming sextuplet lenses
(eq. [32] in Evans \& Witt 2001).
There are only 11 galaxies that fall outside the critical
lines (and one almost on the line). Hence if mass roughly follows light
and there is no significant evolution in shapes  of ellipticals as a
function  of redshift (see Pasquali et al. 2006), then the fraction of galaxies
that can produce more than 4 images will be of the order of
$\sim 1$ per cent. However, the magnification bias (e.g., Turner, Ostriker \&
Gott 1984) for these systems is likely
somewhat higher, bringing the fraction somewhat higher. So far, about
100 gravitational galaxy-scale lenses are known, and only one system shows six images
due to a compact group of galaxies (B1359+154, Rusin et al. 2001).
Currently the number of known lenses ($\sim
100$) is too small to allow a definitive comparison.

Hydrodynamical simulations of galaxies cannot yet simulate
individual elliptical galaxy realistically 
(see, e.g. Meza et al. 2003 for an attempt), let alone the population as a whole.
The statistical results we obtained in this paper 
can serve as a comparison sample for future numerical simulations.
An intermediate, and profitable step is to use collisionless simulations of disk
galaxy mergers to study the isophotal shapes (Naab \& Burkert 2003;
Jesseit, Naab \& Burkert 2005). Naab \& Burkert (2003)
ran 112 merger simulations with mass ratios of 1:1, 2:1, 3:1 and 4:1.
Their 1:1 disk galaxy mergers can reproduce the boxy ellipticals, but 
most of the simulated ellipticals have $a_4/a<0.03$ whatever the mass
ratio (see their fig.~6), so it
will be difficult to reproduce the most discy E/S0s ($a_4/a >= 0.03$), 
about 5.9 per cent of our sample galaxies. Their origin remains
unclear. Perhaps the inclusion of dissipation can remedy this situation.

Weil \& Hernquist (1996) argued that pair and multiple merger remnants 
(e.g., from groups) show differences in both spatial and kinematic properties. 
In particular, they found that multiple merger remnants are more likely
to appear nearly round from many viewing angles. If the multiple mergers
produce more luminous ellipticals, then this would predict a correlation
between the luminosity and the ellipticity.
Fig.~\ref{fig:eSigma} shows that there is indeed some correlation
between the luminosity and dynamical mass versus ellipticity. 
It would be interesting to further explore these correlations with more simulations.

\section*{Acknowledgments}

We thank Drs. Ivo Busko, Neal Jackson, Frederick Kuehn, Zhengyi Shao and
Laura Ferrarese for helpful discussions. Thanks are also due to the anonymous 
referee for constructive comments that improved the paper.
This project is supported by the NSF of China 10333060 and 10273012.
SM acknowledges partial travel support from the Chinese Academy of Sciences
and a visiting professorship from Tianjin Normal University.

  Funding for the SDSS and SDSS-II has been provided by the Alfred P. Sloan Foundation, the Participating Institutions, the National Science Foundation, the U.S. Department of Energy, the National Aeronautics and Space Administration, the Japanese Monbukagakusho, the Max Planck Society, and the Higher Education Funding Council for England. The SDSS Web Site is http://www.sdss.org/.
    The SDSS is managed by the Astrophysical Research Consortium for the Participating Institutions. The Participating Institutions are the American Museum of Natural History, Astrophysical Institute Potsdam, University of Basel, Cambridge University, Case Western Reserve University, University of Chicago, Drexel University, Fermilab, the Institute for Advanced Study, the Japan Participation Group, Johns Hopkins University, the Joint Institute for Nuclear Astrophysics, the Kavli Institute for Particle Astrophysics and Cosmology, the Korean Scientist Group, the Chinese Academy of Sciences (LAMOST), Los Alamos National Laboratory, the Max-Planck-Institute for Astronomy (MPIA), the Max-Planck-Institute for Astrophysics (MPA), New Mexico State University, Ohio State University, University of Pittsburgh, University of Portsmouth, Princeton University, the United States Naval Observatory, and the University of Washington.

\newpage

\appendix
\section{Fitting isophotes with ellipses}

Let us consider the surface brightness on the best-fitting ellipse 
$\rel(\theta)$ given in Eq. (\ref{eq:rtheta}). If $\delta{R}$ is small,
then we have for a fixed $\theta$
\begin{equation}
I(\rel(\theta)) = I(R-\delta R) \approx I(R) - {dI \over dR}\Bigg|_{\rel} \delta R
\end{equation}
Therefore,
\begin{equation}
I(R) = I(\rel(\theta)) + {dI \over dR}\Bigg|_{\rel} \delta R
\end{equation}
As $\langle I(\rel(\theta))\rangle \approx I_0$, we have
\begin{equation}
I(R)
= I_0 + {dI \over dR}\Bigg|_{\rel} \left[a_0+\sum (a_n \cos n \theta + b_n \sin n\theta)\right]
\end{equation}
where we have used the Taylor expansion to the first order in the second step.
By definition, the right hand side of the equation must be equal to 
that of Eq. (\ref{eq:Itheta}). From this, we obtain
\begin{equation}
{dI \over dR}\Bigg|_{\rel} a_n = A_n
\end{equation}
The above equation implies
\begin{equation}
{a_n \over a} = {A_n \over \gamma a}
\label{eq:anAn}
\end{equation}
where a is the semi-major axis length and we have used
\begin{equation}
\gamma = {dI \over dR}\Bigg|_{\rel}
\end{equation}
is the local radial intensity gradient.
The {\tt ellipse} program in IRAF outputs the right hand
side of Eq. (\ref{eq:anAn}), which is identical to the parameter used in
Bender et al. (1989) -- the left hand side of Eq. (\ref{eq:anAn}).


\begin{thebibliography}{}
\bibitem[\protect\citeauthoryear{Becker et al.}{1995}]{Becker95} Becker R. H., White R. L., Helfand D. J., 1995, ApJ, 450, 559 
\bibitem[\protect\citeauthoryear{Bender \& M\"ollenhoff}{1987}]{Bender87}Bender R., M\"ollenhoff C., 1987, A\&A, 177, 71
\bibitem[\protect\citeauthoryear{Bender et al.}{1988}]{Bender88}Bender R., D\"obereiner S., M\"ollenhoff C., 1988, A\&AS,  74, 385
\bibitem[\protect\citeauthoryear{Bender et al.}{1989}]{Bender89}Bender R., Surma P., D\"obereiner S., M\"ollenhoff C., Madejsky R., 1989, A\&A, 217, 35
\bibitem[\protect\citeauthoryear{Bernardi et al.}{2003a}]{Bernardi03a}Bernardi M. et al., 2003, AJ, 125, 1817 (2003a)
\bibitem[\protect\citeauthoryear{Bernardi et al.}{2003b}]{Bernardi03b}Bernardi M. et al., 2003, AJ, 125, 1849 (2003b)
\bibitem[\protect\citeauthoryear{Bertin \& Arnouts}{1996}]{Bertin96}Bertin E., Arnouts S., 1996, A\&AS, 117, 393
\bibitem[\protect\citeauthoryear{Blanton et al.}{2001}]{Blanton01}Blanton M. R. et al., 2001, AJ, 121, 2358
\bibitem[\protect\citeauthoryear{Blanton et al.}{2003}]{Blanton03}Blanton M. R. et al., 2003, AJ, 125, 2348
\bibitem[\protect\citeauthoryear{Blanton et al.}{2005}]{Blanton05}Blanton M. R. et al., 2005, AJ, 129, 2562
\bibitem[\protect\citeauthoryear{Bruzual \& Charlot}{2003}]{Bruzual03}Bruzual G., Charlot S., 2003, MNRAS, 344, 1000
\bibitem[\protect\citeauthoryear{Chang et al.}{2006}]{Chang06}Chang R. X., Gallazzi A., Kauffmann G., Charlot S., Ivezi\'c \v{Z}., Brinchmann J., Heckman T. M., 2006, MNRAS, 366, 717
\bibitem[\protect\citeauthoryear{Condon et al.}{1998}]{Condon98}Condon J. J., Cotton W. D., Greisen E. W., Yin Q. F., Perley R. A., Taylor G. B., Broderick J. J., 1998, AJ, 115, 1693
\bibitem[\protect\citeauthoryear{di Tullio}{1978}]{di Tullio78}di Tullio G., 1978, A\&A, 62, L17
\bibitem[\protect\citeauthoryear{di Tullio}{1979}]{di Tullio79}di Tullio G., 1979, A\&AS., 37, 591
\bibitem[\protect\citeauthoryear{Evans \& Witt}{2001}]{Evans01}Evans N. W., Witt H. J., 2001, MNRAS, 327, 1260
\bibitem[\protect\citeauthoryear{Faber \& Jackson}{1976}]{Faber76}Faber S. M., Jackson R. E., 1976, ApJ, 204, 668
\bibitem[\protect\citeauthoryear{Faber et al.}{1997}]{Faber97} Faber S. M., et al., 1997, AJ, 114, 1771
\bibitem[\protect\citeauthoryear{Fasano \& Vio}{1991}]{Fasano91}Fasano G., Vio R., 1991, MNRAS, 249, 629
\bibitem[\protect\citeauthoryear{Ferrarese et al.}{1994}]{Ferrarese94}Ferrarese L., van den Bosch F. C., Ford H. C., Jaffe W., O'Connell R. W., 1994, AJ, 108, 1598
\bibitem[\protect\citeauthoryear{Jedrzejewski}{1987}]{Jedrzejewski87}Jedrzejewski R. I., 1987, MNRAS, 226, 747
\bibitem[\protect\citeauthoryear{Jesseit et al.}{2005}]{Jesseit05}Jesseit R., Naab T., Burkert A., 2005, MNRAS, 360, 1185
\bibitem[\protect\citeauthoryear{Kauffmann et al.}{2003}]{Kauffmann03}Kauffmann G., et al., 2003, MNRAS, 341, 33
\bibitem[\protect\citeauthoryear{Keeton et al.}{2000}]{Keeton00}Keeton C. R., Mao S., Witt H. J., 2000, ApJ, 537, 697
\bibitem[\protect\citeauthoryear{Khochfar \& Burkert}{2005}]{Khochfar05}Khochfar S., Burkert A., 2005, MNRAS, 359, 1379
\bibitem[\protect\citeauthoryear{Koopmans et al.}{2006}]{Koopmans06}Koopmans L. V. E., Treu T., Bolton A. S., Burles S., Moustakas L. A., 2006, ApJ, in press (astro-ph/0601628)
\bibitem[\protect\citeauthoryear{Kuehn \& Ryden}{2005}]{Kuehn05}Kuehn F., Ryden B. S., 2005, ApJ, 634, 1032
\bibitem[\protect\citeauthoryear{Lambas et al.}{1992}]{Lambas92}Lambas D. G., Maddox S. J., Loveday J., 1992, MNRAS, 258, 404
\bibitem[\protect\citeauthoryear{Lauer}{1985}]{Lauer85}Lauer T. R., 1985, MNRAS, 216, 429
\bibitem[\protect\citeauthoryear{Lauer et al.}{1995}]{Lauer95}Lauer T. R. et al., 1995, AJ, 110, 2622
\bibitem[\protect\citeauthoryear{Lauer et al.}{2005}]{Lauer05}Lauer T. R. et al., 2005, AJ, 129, 2138
\bibitem[\protect\citeauthoryear{Meza et al.}{2003}]{Meza03}Meza A., Navarro J. F., Steinmetz M., Eke V. R., 2003, ApJ, 590, 619
\bibitem[\protect\citeauthoryear{Mo et al.}{1998}]{Mo98}Mo H. J., Mao S., White S. D. M., 1998, MNRAS, 295, 319
\bibitem[\protect\citeauthoryear{Naab \& Burkert}{2003}]{Naab03} Naab T., Burkert A., 2003, ApJ, 597, 893
\bibitem[\protect\citeauthoryear{Naab et al.}{1999}]{Naab99}Naab T., Burkert A., Hernquist L., 1999, ApJ, 523, L133
\bibitem[\protect\citeauthoryear{Naab et al.}{2006}]{Naab06}Naab T., Khochfar S., Burkert A., 2006, ApJ, 636, L81
\bibitem[\protect\citeauthoryear{Pasquali et al.}{2006}]{Pasquali06} Pasquali A. et al., 2006, ApJ, 636, 115
\bibitem[\protect\citeauthoryear{Pellegrini}{1999}]{Pellegrini99}Pellegrini S., 1999, A\&A, 351, 487
\bibitem[\protect\citeauthoryear{Pellegrini}{2005}]{Pellegrini05}Pellegrini S., 2005, MNRAS, 364, 169
\bibitem[\protect\citeauthoryear{Reda et al.}{2004}]{Reda04}Reda F. M., Forbes D. A., Beasley M. A., O'Sullivan E. J., Goudfrooij P., 2004, MNRAS, 354, 851
\bibitem[\protect\citeauthoryear{Rest et al.}{2001}]{Rest01} Rest A., van den Bosch F. C., Jaffe W., Tran H., Tsvetanov Z., Ford H. C., Davies J., Schafer J., 2001, AJ, 121, 2431
\bibitem[\protect\citeauthoryear{Rusin et al.}{2001}]{Rusin01}Rusin D. et al., 2001, ApJ, 557, 594
\bibitem[\protect\citeauthoryear{Ryden et al.}{2001}]{Ryden01}Ryden B. S., Fobes D. A., Terlevich A. I., 2001, MNRAS, 326, 1141
\bibitem[\protect\citeauthoryear{S\'ersic}{1968}]{sersic68}S\'ersic J. L., 1968, Atlas de Galaxias Australes Cordoba: Observatorio Astronomico
\bibitem[\protect\citeauthoryear{Sheth et al.}{2003}]{Sheth03}Sheth R. K. et al., 2003, ApJ, 594, 225
\bibitem[\protect\citeauthoryear{Shioya \& Taniguchi}{1993}]{Shioya93}Shioya Y., Taniguchi Y., 1993, PASJ, 45, L39
\bibitem[\protect\citeauthoryear{Smith et al.}{2002}]{Smith02}Smith J. A. et al., 2002, AJ, 123, 2121
\bibitem[\protect\citeauthoryear{Stoughton et al.}{2002}]{Stoughton02}Stoughton C. et al., 2002, AJ, 123, 485 
\bibitem[\protect\citeauthoryear{Turner et al.}{1984}]{Turner84}Turner E. L., Ostriker J. P., Gott III J. R., 1984, ApJ, 284, 1 
\bibitem[\protect\citeauthoryear{van den Bosch et al.}{1994}]{Bosch94}van den Bosch F. C., Ferrarese L., Jaffe W., Ford H. C., O'Connell R. W., 1994, AJ, 108, 1579
\bibitem[\protect\citeauthoryear{Vincent \& Ryden}{2005}]{Vincent05} Vincent R. A., Ryden B. S., 2005, ApJ, 623, 137
\bibitem[\protect\citeauthoryear{Weil \& Hernquist}{1996}]{Weil96} Weil M. L., Hernquist L., 1996, ApJ, 460, 101
\bibitem[\protect\citeauthoryear{Wu et al.}{2005}]{Wu05} Wu H., Shao Z. Y., Mo H. J., Xia X. Y., Deng Z. G., 2005, ApJ, 622, 244
\bibitem[\protect\citeauthoryear{Yoo et al.}{2006}]{Yoo06}Yoo J., Kochanek C. S., Falco E. E., McLeod B. A., 2006, ApJ, 642, 22
\end{thebibliography}
\end{document}